\documentclass[acmtog, authorversion]{acmart}
\newcommand{\final}{1}

\usepackage[english]{babel}
\usepackage{amsmath}
\usepackage{bm}

\usepackage[toc,page,titletoc]{appendix}
\usepackage[capitalise,nameinlink]{cleveref}
\crefname{supp}{Supplement}{Supplements}

\usepackage{caption}
\usepackage{subfig}
\usepackage{wrapfig}
\usepackage{mathtools}
\usepackage{ifthen}
\usepackage[most]{tcolorbox}
\usepackage[export]{adjustbox}
\usepackage{algorithm}
\usepackage{algpseudocode}
\usepackage{longtable}

\providecommand\BibTeX{{\normalfont B{\scshape i\kern-0.0em b}\TeX}}

\definecolor{lightgreen}{rgb}{0.56, 0.93, 0.56}
\definecolor{moonstoneblue}{rgb}{0.45, 0.66, 0.76}
\definecolor{randColor}{rgb}{0.9, 0.51, 0.1}

\newcommand{\warning}[1]{{\it\color{red} #1}}
\newcommand{\note}[1]{{\it\color{blue} #1}}
\newcommand{\nothing}[1]{}
\usepackage[normalem]{ulem}
\newcommand{\revision}[2]{#2}
\newcommand{\kenny}[2]{\note{\color{teal}Kenny #1: #2}}
\newcommand{\monde}[2]{\note{Monde #1: #2}}
\newcommand{\qisun}[1]{{\it\color{randColor}Qi: #1}}

\newcommand{\fixme}[1]{{\textcolor{red}{\textit{#1}}}}

\ifthenelse{\equal{\final}{1}} {
    \renewcommand{\warning}[1]{}
    \renewcommand{\note}[1]{}
    \renewcommand{\monde}[1]{}
    \renewcommand{\qisun}[1]{}
    \renewcommand{\kenny}[1]{}
    \renewcommand{\fixme}[1]{}
}

\newcommand{\Caption}[2]{\caption[#1]{{\em #1} #2}}

\renewcommand{\vec}[1]{\mathbf{#1}}
\usepackage{tikz}

\newcommand{\testVec}{\vec{t}}
\newcommand{\testCoordinate}{t}
\newcommand{\adaptationVec}{\vec{b}}
\newcommand{\adaptationCoordinate}{b}
\newcommand{\colorContrastVec}{\bm{\kappa}}
\newcommand{\colorContrastCoordinate}{\kappa}

\newcommand{\ellipse}{\mathcal{E}}
\newcommand{\colorVec}{\vec{x}}
\newcommand{\colorCoordinate}{x}
\newcommand{\normalizedEllipseVec}{\bm{\alpha}}
\newcommand{\normalizedEllipseCoordinate}{\alpha}

\newcommand{\real}{\mathbb{R}}

\newcommand{\neuralNetwork}{\Phi}
\newcommand{\perceptualModel}{\bm{\hat{\alpha}}}
\newcommand{\eccentricity}{e}
\DeclareMathOperator*{\argmin}{arg\,min}

\newcommand{\rbfNodes}{N}
\newcommand{\rbfWeight}{\bm\lambda}
\newcommand{\rbfBias}{\bm\nu}
\newcommand{\rbfCentre}{\mathbf{c}}
\newcommand{\rbfStd}{\sigma}
\newcommand{\gaussianBasis}{\rho}
\newcommand{\ellipseMax}{\bm{\eta}}

\newcommand{\ellipseVec}{\vec{a}}
\newcommand{\ellipseCoordinate}{a}

\newcommand{\power}{\mathcal{P}}
\newcommand{\powerCirc}{p_{circ}}
\newcommand{\powerVec}{\mathbf{p}}
\newcommand{\powerCoordinate}{p}
\newcommand{\colorSpaceTransform}{M}

\newcommand{\conditionUniform}{\mathbf{LUM}}
\newcommand{\conditionOurs}{\mathbf{OUR}}

\graphicspath{
{figures/}
}

\setcopyright{acmcopyright}
\copyrightyear{2022}
\acmJournal{TOG}
\acmYear{2022}
\acmVolume{41}
\acmNumber{6}
\acmArticle{210}
\acmMonth{12}
\acmDOI{10.1145/3550454.3555473}
\acmSubmissionID{papers\_307}
\begin{document}

\title{Color-Perception-Guided Display Power Reduction for Virtual Reality}

\author{Budmonde Duinkharjav}
\authornote{equal contribution.}
\email{budmonde@gmail.com}
\orcid{0000-0001-7133-3273}
\author{Kenneth Chen}
\authornotemark[1]
\email{kc4906@nyu.edu}
\orcid{0000-0002-8095-4407}
\affiliation{%
    \institution{New York University}
    \country{USA}
}

\author{Abhishek Tyagi}
\email{atyagi2@ur.rochester.edu}
\orcid{0000-0003-0458-0172}
\affiliation{%
    \institution{University of Rochester}
    \country{USA}
}

\author{Jiayi He}
\email{jhe36@u.rochester.edu}
\orcid{0000-0002-9014-0941}
\affiliation{%
    \institution{University of Rochester}
    \country{USA}
}

\author{Yuhao Zhu}
\authornote{corresponding authors.}
\email{yzhu@rochester.edu}
\orcid{0000-0002-2802-0578}
\affiliation{%
    \institution{University of Rochester}
    \country{USA}
}

\author{Qi Sun}
\authornotemark[2]
\email{qisun@nyu.edu}
\orcid{0000-0002-3094-5844}
\affiliation{%
    \institution{New York University}
    \country{USA}
}



\begin{teaserfigure}
    \centering
    \subfloat[original (left) vs. our power-optimized frame that preserves perceptual fidelity (right)]{%
      \includegraphics[width=0.6\linewidth,valign=t]{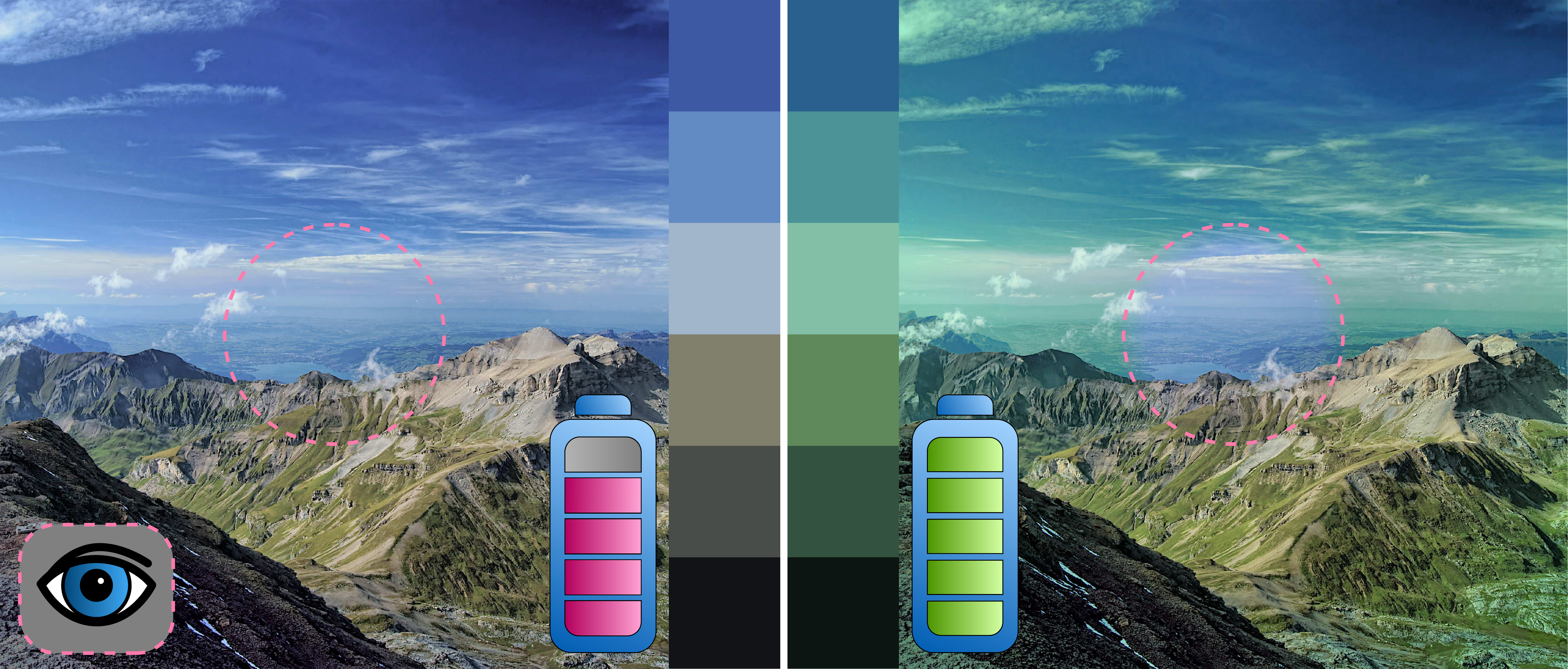}
      \vphantom{\includegraphics[width=0.4\linewidth,valign=t]{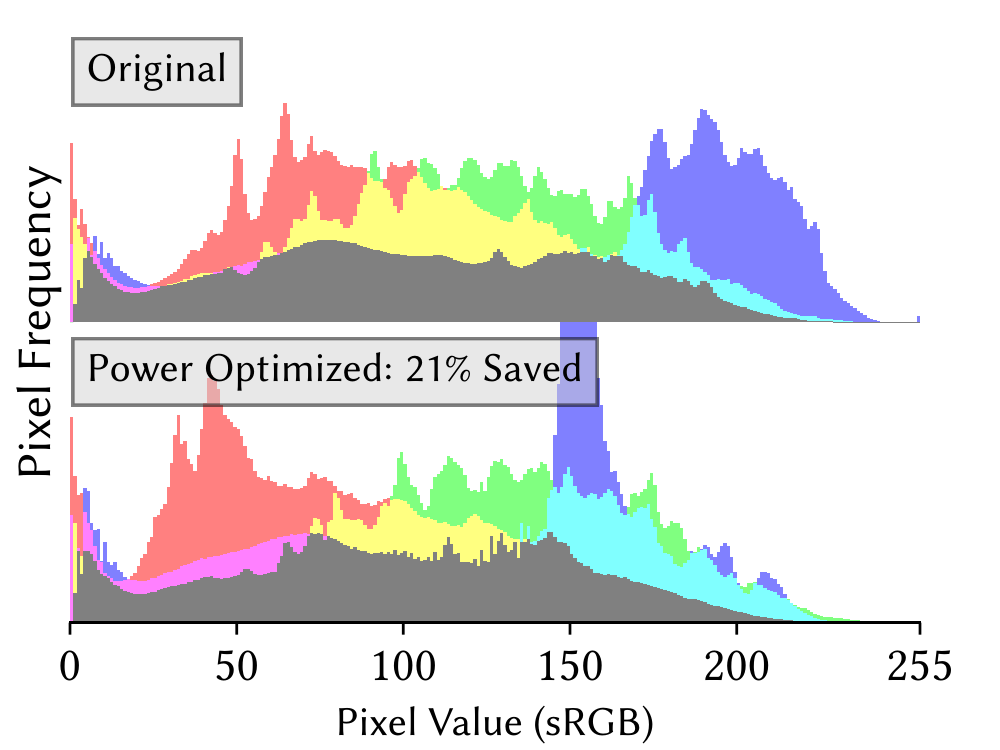}}
      \label{fig:teaser:main}
    }%
    \subfloat[image statistics]{%
      \includegraphics[width=0.4\linewidth,valign=t]{figures/teaser/teaser_histogram.pdf}
      \label{fig:teaser:analysis}
    }%
    \Caption{%
        Illustration of our gaze-contingent and color-perception-aware display power saving model.%
    }{%
        We present a perceptually-guided, real-time, and closed-form model for minimizing the power consumption of untethered VR displays while preserving visual fidelity.
        We apply a gaze-contingent shader onto the original frame (\subref{fig:teaser:main} left) to produce a more power efficient frame (\subref{fig:teaser:main} right) while preserving the luminance level and perceptual fidelity during active viewing. The dashed circles indicate the user's gaze.
        Our method is jointly motivated by prior literature revealing that: i) the power cost of displaying different colors on LEDs may vary significantly, even if the luminance levels remain unchanged \cite{dong2011chameleon}; ii) human  color sensitivity decreases in peripheral \cite{Hansen:2009:ColorVsEccentricity} and active vision \cite{Cohen:2020:TLC}.
        The color palette of the original frame is modulated using our peripheral filter and is shown in \subref{fig:teaser:main} for a visual comparison.
        While the color palettes appear different when gazed upon, an observer cannot discriminate between them when shown in their periphery.
        \subref{fig:teaser:analysis} visualizes how our model shifts the image's chromatic histograms to minimize the physically-measured power consumption. 
        The blue LEDs consume more energy than the red/green in our experiment display panel.
        Image credits to Tim Caynes \copyright\ 2012.
    }%
    \Description{Illustration of our model performance.}
  \label{fig:teaser}
\end{teaserfigure}

\begin{abstract}
Battery life is an increasingly urgent challenge for today's untethered VR and AR devices. However, the power efficiency of head-mounted displays is naturally at odds with growing computational requirements driven by better resolution, refresh rate, and dynamic ranges, all of which reduce the sustained usage time of untethered AR/VR devices. For instance, the Oculus Quest 2, under a fully-charged battery, can sustain only 2 to 3 hours of operation time. Prior display power reduction techniques mostly target smartphone displays. Directly applying smartphone display power reduction techniques, however, degrades the visual perception in AR/VR with noticeable artifacts. For instance, the ``power-saving mode'' on smartphones \emph{uniformly} lowers the pixel luminance across the display and, as a result, presents an overall darkened visual perception to users if directly applied to VR content.

Our key insight is that VR display power reduction must be cognizant of the gaze-contingent nature of high field-of-view VR displays. To that end, we present a gaze-contingent system that, without degrading luminance, minimizes the display power consumption while preserving high visual fidelity when users actively view immersive video sequences. This is enabled by constructing 1) a gaze-contingent color discrimination model through psychophysical studies, and 2) a display power model (with respect to pixel color) through real-device measurements. Critically, due to the careful design decisions made in constructing the two models, our algorithm is cast as a constrained optimization problem with a \emph{closed-form} solution, which can be implemented as a real-time, image-space shader. We evaluate our system using a series of psychophysical studies and large-scale analyses on natural images. Experiment results show that our system reduces the display power by as much as $24\%$ ($14\%$ on average) with little to no perceptual fidelity degradation.
\end{abstract}

\maketitle

\section{Introduction}
\label{sec:introduction}

Virtual and Augmented Reality (VR and AR) devices are increasingly becoming untethered for portability, outdoor usage, and unrestricted locomotion to enable ultimate immersion. 
At the same time, the demands for higher resolution, framerate, and dynamic range are steadily increasing, which is directly at odds with the limited energy capacity of untethered AR/VR devices.
For example, when fully charged, both the Oculus Quest 2 and Hololens 2 can actively run only for 2-3 hours \cite{holoBattery}. 
Since the total energy capacity increases only marginally because ``there is no Moore's law for batteries'' \cite{schlachter2013no}, power consumption has become a primary concern in the design process of AR/VR devices \cite{wang2016real,zhang2021powernet,debattista2018frame}.

This paper specifically focuses on reducing the display power. In our measurement of HTC Vive Pro Eye and Oculus Quest 2, the display consumes as much as half of the total power consumption by comparing the power when the display is on vs. off. The results are consistent with data reported in other measurement studies \cite{leng2019energy, Yan:2018:EEA}.
Display power will only become more important in the cloud rendering paradigm, where the computation is offloaded to the cloud, heightening the contribution of display to the total device power.

Conventional display power optimizations are geared toward smartphones, which, when directly applied to VR devices, lead to significant visual quality degradation.
This is because smartphone display optimizations are fundamentally gaze-\textit{agnostic}, rightly so because smartphone displays have very narrow field-of-view. These optimizations either modulate pixels \textit{uniformly} across the display \cite{shye2009into, Yan:2018:EEA} or are purely based on the content (e.g., UI elements) \cite{dong2009power, dong2011chameleon, ranganathan2006energy}.
Classic gaze-contingent optimizations in AR/VR such as foveated rendering, while reducing the rendering load \cite{Krajancich:2020:spatiotemp_model, Patney:2016:TFR}, do not (directly) reduce the display power.

We present a gaze-contingent rendering approach that reduces the power consumption of untethered VR displays by as much as $24\%$ while preserving visual quality during active viewing.
We achieve this by only modulating the chromaticity of the display output without changing luminance.

Our method is jointly motivated by hardware research that revealed the variation of power consumption of displaying different colors on LEDs \cite{dong2011chameleon}, as well as the recently discovered limitations of human peripheral color perception during active vision \cite{Cohen:2020:TLC}.
That is, given an original frame such as in a 360 video, we seek a computational model that guides a gaze-contingent color ``shift'' that (1) requires the minimal power cost, and (2) preserves the perceived fidelity.

To accomplish this, we conducted two pilot studies.
First, we quantitatively model how our color sensitivity degrades with higher retinal eccentricities.
Second, we physically measure the LED display power consumption as a function of the displayed color.
Given the perceptual and the power model, our system performs a constrained optimization that identifies, for each pixel, an alternative color that minimizes the power consumption while maintaining the same perceptual quality. Critically, the optimization problem has a \textit{closed-form} solution because of the judicious design decisions we made in constructing the perceptual and power models.
As a result, our perception-perserving color modulation can be implemented as a real-time shader.

We validate our method with both subjective studies on panoramic videos, as well as an objective analysis on large-scale natural image data.
We demonstrate the model's effectiveness in display power reduction and perceptual fidelity preservation, relative to an alternative luminance-based ``power saver''.
Our objective analysis concludes that this model shows generalizability to a large variety of natural scenes and save, on average, $14\%$ power.
In summary, our main contributions include:
\begin{itemize}
  \item a psychophysical study and data that measure human color discrimination sensitivity at various retinal eccentricities and reference colors in the Derrington-Krauskopf-Lennie (\emph{DKL}) color-space;
  \item a physical system and data that measure the power consumption of VR-alike stereoscopic displays as a function of displayed color;
  \item a closed-form formulation that suggests the optimal (in terms of lowest display power cost without compromised visual fidelity) per-pixel chromaticity modulation by leveraging two learned models from our two aforementioned datasets;
  \item a real-time shader for gaze-tracked VR headsets and natural content viewing applications, as well as a demonstration of its general benefits with a large-scale analysis.
\end{itemize}

We provide the source code for our model regression and shader implementations at \url{www.github.com/NYU-ICL/vr-power-saver}.

\section{Related Work}
\label{sec:related}
\subsection{Energy-Aware Graphics and Display}

The graphics rendering pipeline requires heavy computation to execute in real-time. Reliably maintaining the performance requirements of these applications consumes considerable power. As such, energy-aware methods have been developed to minimize power while maintaining rendering quality. Most prior work in the graphics literature, however, focuses on reducing the rendering power \cite{wang2016real,zhang2018fly,zhang2021powernet,debattista2018frame}.

Complementary to prior work on reducing the rendering power, our work reduces the \textit{display} power --- by modulating the display color while preserving perceptual fidelity. 
The relationship between display power and color is studied in the mobile computing community, mostly in the context of smartphones. Dash and Hu \shortcite{Dash2021}, Dong et al. \shortcite{dong2009power}, and Dong and Zhong \shortcite{dong2011chameleon} model the power consumption of smartphone displays with respect to color. Miller et al. \shortcite{miller2006p} discusses how various hardware design optimizations can affect the power modeling accuracy.

Other work focuses on reducing the display power by modulating the brightness/luminance of the display rather than color, which our work focuses on. 
For instance, Shye et al. \shortcite{shye2009into} gradually dims the smartphone display after a user stares at the display for an extended period of time. Yan et al. \shortcite{Yan:2018:EEA} pushes the empirical approach in Shye et al. \shortcite{shye2009into} a step further, and uses physiological data to derive a quantitative method to adaptively reduce the display luminance.
Our work on color modulation is orthogonal and complementary to luminance-modulating techniques.
We show that significant power saving is readily obtainable by adjusting only color; combining color and luminance modulation would conceivably lead to higher power savings, which we leave to future work (see \Cref{sec:future-work}).

Another orthogonal line of work is to reduce OLED power via better hardware design. Shin et al. \shortcite{shin2013dynamic} propose to dynamically scale the display supply voltage, coupled with image-space color transformation, to minimize the display power while maintaining perceptual color similarity (as quantified by the CIELAB $\Delta E^\ast$ metric). Boroson et al. \shortcite{boroson2009oled}, Miller et al. \shortcite{miller2008color} and Miller et al. \shortcite{miller2009oled} propose four-color OLED structures, where a fourth sub-pixel has a higher power efficiency (luminous efficacy) than one of the RGB sub-pixels. The fourth, higher-efficiency sub-pixel 
allows many colors that usually dominate an image (e.g., neutral or saturated colors) to be produced with less contribution from the blue and red sub-pixels, which have low power efficiency. These prior works, however, do not consider the gaze-contingent color perception in VR.

\subsection{Perceptually-Aware Immersive Rendering}
Perceptual graphics studies hinge on the idea that the human visual system's ability to receive and process light signals deteriorates for stimuli located at higher retinal eccentricities, due to the significantly denser distribution of retinal cone cells in the fovea \cite{Song:2011:ConeDensity}.
Studies have taken advantage of this fact to create perceptually-aware models of rendering imagery \cite{Duchowski:2004}. 
Foveated rendering takes advantage of the drop in visual acuity as retinal eccentricity increases by reducing image quality in the periphery, ultimately improving performance while maintaining perceptual quality \cite{Patney:2016:TFR,Sun:2017:PGF,guenter2012foveated,walton2021beyond}. 
Similar perceptually-aware gaze-contingent algorithms have been proposed to save spatio-temporal data bandwidth  \cite{Krajancich:2020:spatiotemp_model,kaplanyan2019deepfovea}.
However, other than modulating user-perceived color \cite{mauderer2016gaze} or remapping peripheral luminance and color for power saving \cite{li_patent_2022}, there has been little work leveraging gaze-contingent color perception to optimize real-time rendering.
That could be mainly because the modulated color does not introduce performance acceleration within the graphics pipeline.

\subsection{Visual Perception of Displayed Color}
\label{sec:prior:perception}
Color and the corresponding human perception is a long-standing and extensive research topic in computer graphics \cite{rhyne2018color}. Here, we discuss prior research that understands perceptual effects on display color.

First, perception of color is studied from psychophysical and even physiological perspectives.
Early color perception research was largely based on psychophysical color matching experiments, which resulted in the \emph{CIE 1931 RGB} \cite{wright1929re,guild1931colorimetric} and \emph{CIE 1931 XYZ} color spaces \cite{fairman1997cie}. The \emph{CIE XYZ} color space has since become the cornerstone of modern color research, because it presents a hardware/device-independent way of quantifying colors.
In fact, the first measurements of human color-discriminative thresholds, widely known as the MacAdam ellipses were presented in the \emph{xy} chromaticity plot \cite{Macadam:1942:Ellipses}.

The \emph{sRGB} color space widely used in graphics, display, and vision fields today is derived from the \emph{XYZ} color space.
The \emph{sRGB} color space, however, is
rarely directly used in color perception studies, because \emph{sRGB} is a device-dependent color space with non-uniform quantization and has a gamut smaller than the gamut of the human visual system.
Color perception studies usually operate on some form of physiologically-based \emph{LMS} color spaces, which are developed from measurements of the surgically removed cone cells \cite{bowmaker1980visual,dartnall1983human} or from color defective vision~\cite{stockman2000spectral}.
Cone color spaces provide the ability to make ``feed-forward'' reasoning about our visual perception system.

One such physiologically-based study of color-discrimi\-native thresholds was established in \cite{Krauskopf:1992:DiscriminationAdaptation}.
Similar to the color-opponent theory, upon which CIELAB was based on \cite{schiller1990color}, neural recordings within the visual cortex confirmed the existence of a cone-opponent mechanism where signals from the $L$, $M$, and $S$ cones are compared in a feed-forward fashion \cite{DeValois:1966:LGNanalysis}.
The \emph{DKL} color space \cite{Derrington:1984:DKL} leverages this cone-opponent mechanism to derive a physiologically-relevant and perceptually uniform color space. Our perceptual study is based on the \emph{DKL} color space.

Second, human color perception exhibits eccentricity effects. Similar to our resolution and depth \cite{sun2020eccentricity} acuity, color sensitivity also decreases as retinal eccentricity increases \cite{Hansen:2008:ColorNatural,Hansen:2009:ColorVsEccentricity,Cohen:2020:TLC}. 
For instance, given a reference color, our discrimination thresholds are observed to have the shape of an ellipse in \emph{DKL} space, and this region of sub-threshold colors increases significantly as the retinal eccentricity of stimuli increase ($\sim4.5\times$ larger ellipse radii at $50^\circ$ compared to $5^\circ$) \cite{Hansen:2009:ColorVsEccentricity}.
In AR, the interference between virtual and physical colors raises new challenges in correctly aligning perceived color \cite{murdoch2015towards,hassani2016color,zhang2021color}.
However, to our knowledge, there is no computational model that numerically predicts the sensitivity given an eccentricity and reference color.

Third, chromaticity and luminance have different temporal roles and sensitivities \cite{hermann2021temporal}. 
The eccentricity effects of luminance contrast have been leveraged to perform gaze-contingent rendering \cite{tursun2019luminance} and measure spatio-temporal video quality, compared with a reference \cite{mantiuk2021fovvideovdp}.
Chromaticity-based sensitivity has been leveraged to encode high dynamic range displays \cite{kim2021color}.
However, perhaps because modulating colors does not play a role in accelerating performance, the eccentricity effect has not been investigated for advancing VR/AR systems.

Finally, our color sensitivity is also correlated to human status and task nature.
For instance, the color sensitivity during fixation shifting (a.k.a. saccade) uniformly and significantly decreases \cite{braun2017visual}.
Prior literature showed notably lower sensitivity in discrimination than detection tasks \cite{vingrys1998color}. More recently, studies by Cohen et al. \shortcite{Cohen:2020:TLC} revealed the remarkably further decreased color sensitivity during active and natural viewing tasks: in those conditions, even desaturating peripheral images are imperceptible by viewers. Those studies all exhibit the complex nature of color perception, which is still undergoing active scientific discoveries. 
In this research, we focus on \emph{active viewing} scenarios, which are representative conditions of VR/AR applications such as gaming and video-watching.

\begin{figure*}
    \centering
    \subfloat[task protocol]{
      \includegraphics[width=0.32\linewidth,valign=t]{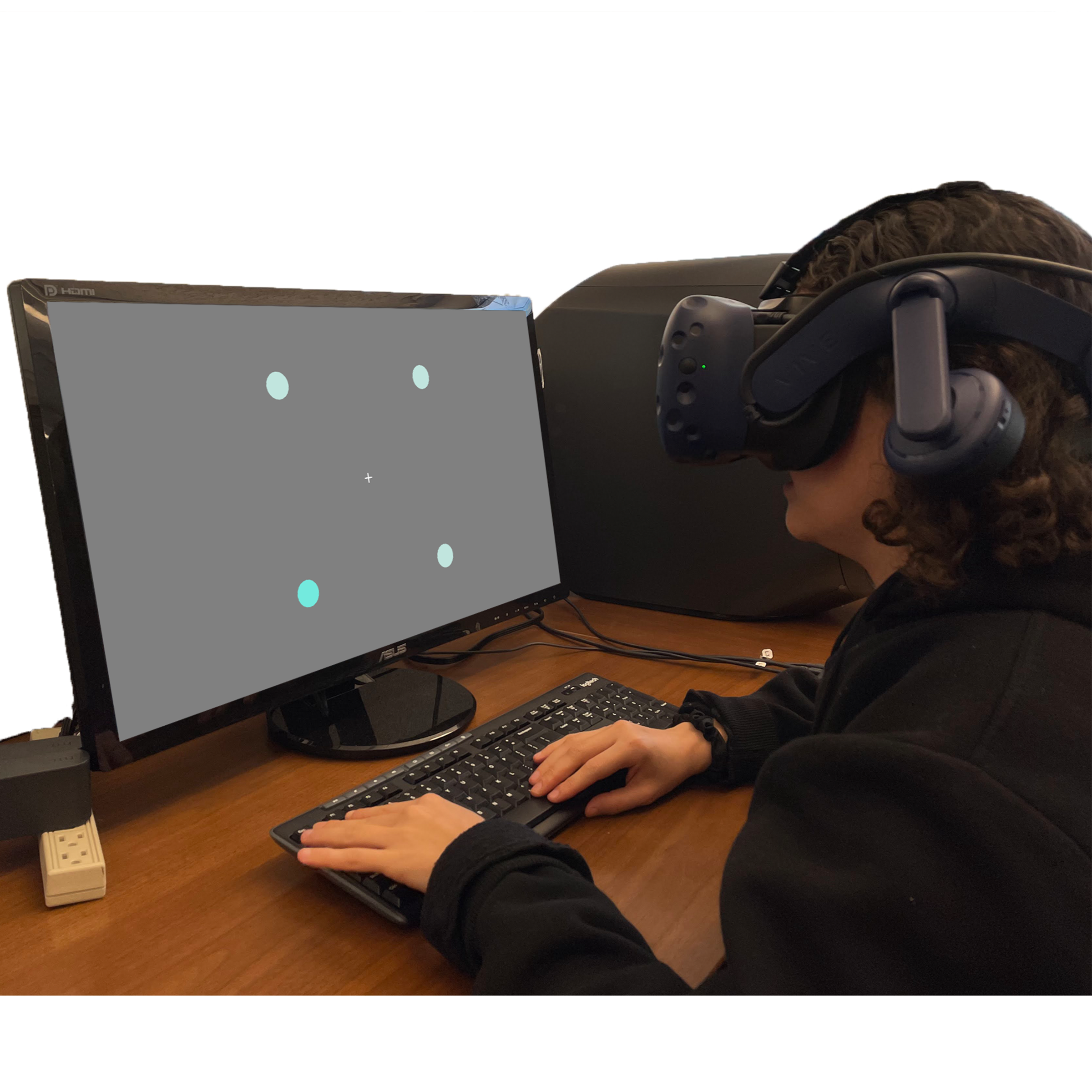}
      \vphantom{\includegraphics[width=0.32\linewidth,valign=t]{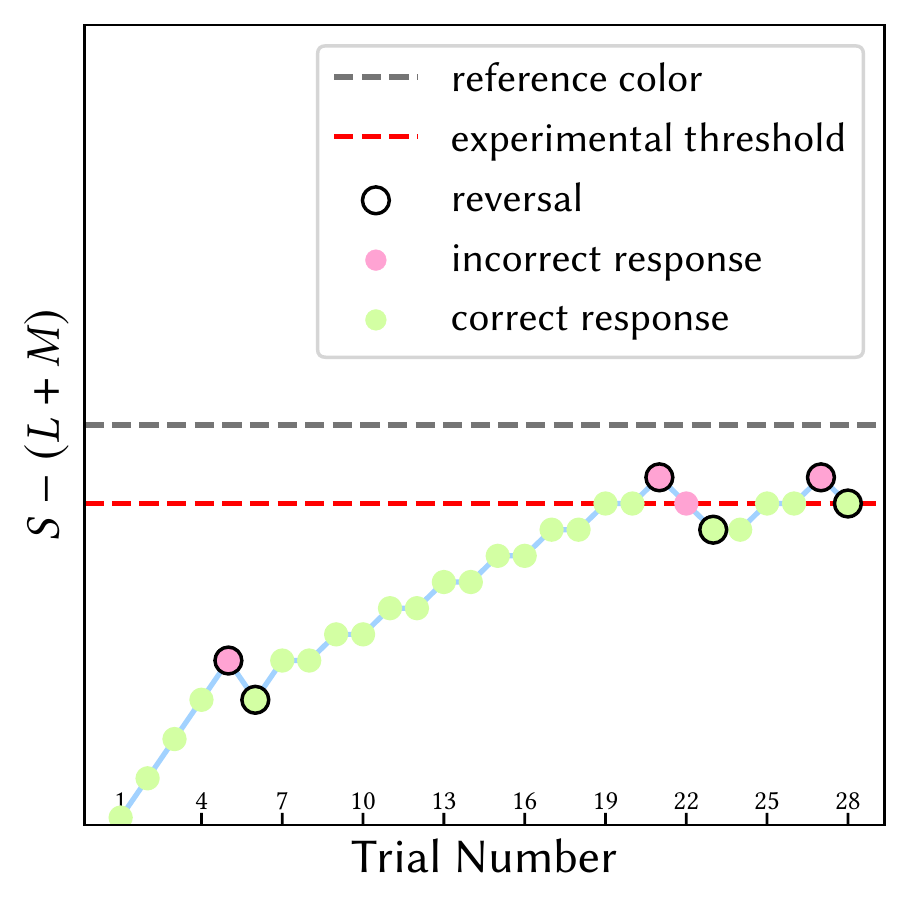}}
      \label{fig:perceptual:task-protocol}
    }%
    \subfloat[example staircase procedure]{
      \includegraphics[width=0.32\linewidth,valign=t]{figures/perceptual-study/staircase_LSM_rw_2.pdf}
      \label{fig:perceptual:staircase-procedure}
    }%
    \subfloat[color discrimination thresholds]{
      \includegraphics[width=0.32\linewidth,valign=t]{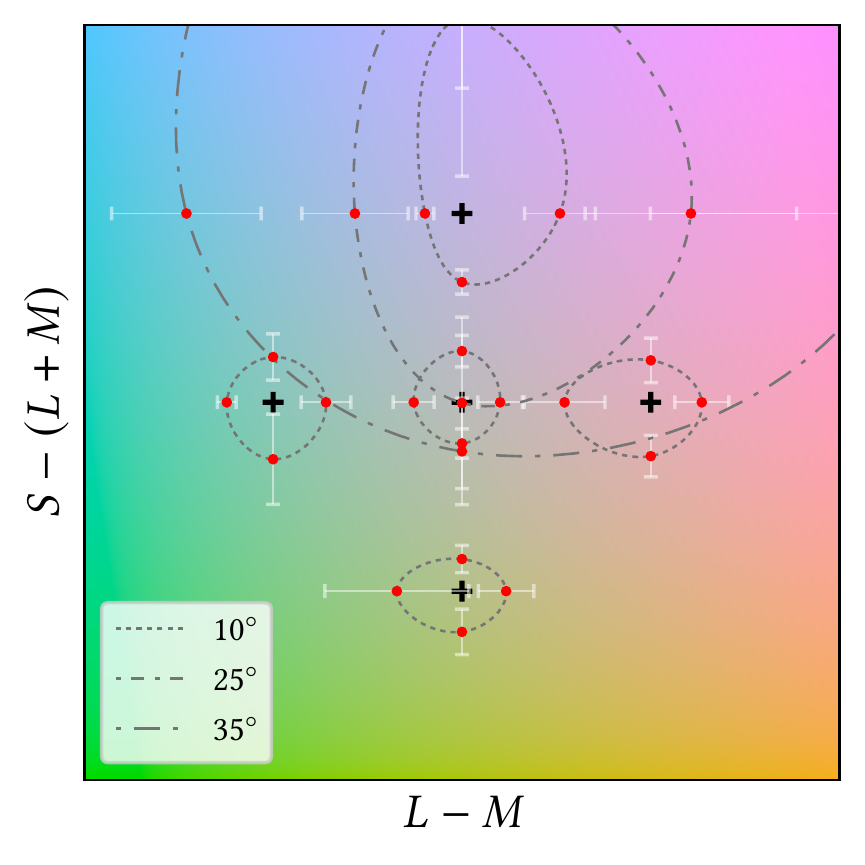}
      \label{fig:perceptual:thresholds}
    }%
    \Caption{
       Pilot study for color discrimination threshold.
    }{
        The $S-(L+M)$ and $L-M$ axes are axes in the \emph{DKL} color-space, which  compares the difference between $S$ vs $L+M$ and $L$ vs $M$ cone activations \cite{Derrington:1984:DKL}.
        \subref{fig:perceptual:task-protocol}
        Participants were instructed to fix their gaze at the center of a VR headset display and to distinguish the ``calibration'' disk out of the four shown in their periphery.
        The other three ``reference'' disks were colored with the reference color, while the calibration disk's color is adjusted via a 1-up-2-down staircase procedure by the participant.
        \subref{fig:perceptual:staircase-procedure}
        The staircase procedure aims to narrow down the peripheral color-discrimination threshold of the participant.
        As the participant completes a series of 4AFC tasks as depicted in \subref{fig:perceptual:task-protocol}, the staircase procedure automatically adjusts the color of the calibration disk to make it harder if the participant identified the calibration disk correctly and vice-versa.
        We continue each trial until a total of $6$ reversals (denoted by outlined points) occur or $50$ trials are completed, whichever occurs first.
        In this example, the experiment continues until the threshold in the $S-(L+M)$ axis is converged.
        \subref{fig:perceptual:thresholds}
        The black crosses indicate our $5$ sampled reference/pedestal colors in \emph{DKL} space. 
        The experimental thresholds are displayed with red dots as mean, and white bars as $75\%$ confidence intervals.
        For the other two sampled eccentricities ($25^\circ$/$35^\circ$), we only plot the results from one reference color, at 12 o'clock in the figure, to avoid visual clutter.
        We plot splines (dashed gray lines) for ease of visualization only.
    }
    \Description{Perceptual User Study for Color Discrimination Thresholds}
    \label{fig:perceptual}
\end{figure*}

\section{Pilot Study: Eccentricity Effects on Color Perception}
\label{sec:perceptual-study}
We aim to exploit how human perception of color varies across the visual field, so that we can adjust the appearance of visual stimuli in our peripheral vision in an advantageous way.
Hansen et al.~\shortcite{Hansen:2009:ColorVsEccentricity} showed that while our ability to discriminate colors significantly deteriorates at high retinal eccentricities, we still maintain some ability to discriminate colors at eccentricities as high as $45^\circ$.
Drawing inspiration from this work, we designed and performed a psychophysical study on the perceptual \emph{discrimination} thresholds of colors, given various reference colors ($5$ total) and retinal eccentricities (from $10^\circ$ to $35^\circ$).
The experimental data later transforms to a computational model in \Cref{sec:method:color}.

\paragraph{Setup}
We perform our study with the HTC Vive Pro Eye head-mounted display as shown in \Cref{fig:perceptual:task-protocol}.
Participants remained seated during the duration of the study, and interacted with the user study software via the keyboard.

\paragraph{Participants}
We recruited 5 participants (ages 20-32, 2 female) for a series of  four-alternative forced choice (4AFC) staircase experiments (similar to \cite{Hansen:2009:ColorVsEccentricity}) to determine the discrimination thresholds.
All participants had normal or corrected-to-normal vision and exhibited no color perception deficits as tested by the Ishihara pseudo-isochromatic plates.
In this pilot study, we chose 5 participants due to the long duration of our staircase experiment. This is also practiced for similar threshold-determination psychophysical experiments \cite{sun2020eccentricity, Krajancich:2020:spatiotemp_model}.
All experiments were approved by an ethics committee and all participants' data was de-identified.

\paragraph{Stimuli}
As shown in \Cref{fig:perceptual:task-protocol}, the stimuli were four colored disks (with a diameter of 5 degrees).
They were rendered simultaneously on top of a neutral gray background (i.e., $[0.5, 0.5, 0.5]$ in the linear \emph{sRGB} space, or $71.5$~cd$/\text{m}^2$).
The azimuth position of the disks remained constant throughout the entire study, located at $45^\circ$, $135^\circ$, $225^\circ$, and $315^\circ$ (i.e. the four diagonals in the participant's visual field), while the radial position (i.e., the retinal eccentricity) varied across sequences to be either $10^\circ$, $25^\circ$, or $35^\circ$.
Three of the disks have the same ``reference'' color, and the fourth has a ``calibration'' color which changes throughout a sequence of trials.
The space of colors that the disks can obtain is visualized as a color-space in \Cref{fig:perceptual:thresholds}.
The luminance of all disks is maintained at the same level as the background's luminance.

\paragraph{Tasks}
The task was an 1-up-2-down 4AFC staircase procedure. 
Specifically, the study was conducted in a single session split into $60$ sequences ($=5$ reference colors $\times$ $3$ eccentricities $\times$ $4$ color space dimensions, as specified below) of trials.
Each sequence was a staircase procedure meant to determine a participant's discriminative threshold of the calibration disk, which was colored with subtly different chromaticities.
Each sequence may contain a variable number of trials depending on the staircase convergence.

Each trial was a single 4AFC task where the participants were instructed to identify which one of the 4 color disks appeared different. The participant was instructed to fix their gaze on a white crosshair at the center of the screen for the duration that the stimuli were shown.
We used eye tracking to ensure participants maintain their gaze at the central crosshair. We automatically rejected a trial if the user’s gaze moves beyond 3$^\circ$ eccentricity, randomized the trial order again, and notified them.
At the start of each trial of a sequence, we shuffle the four colored disks, and display them for $500$ms (the same stimulus duration used in prior color discrimination literature \cite{Hansen:2009:ColorVsEccentricity}).
Once the stimuli disappear, we prompt the participant to identify and select the disk with the calibration color, using the keyboard.
Depending on their answer, the calibration disk's color was made easier or harder to discriminate in the subsequent trial by adjusting the chromaticity of the calibration disk to be closer/farther from the reference color while preserving its luminance.
After $6$ reversals of this staircase procedure (or a maximum of $50$ trials), the sequence terminates, and the next sequence begins.
We visualize the progression of an example staircase-procedure in \Cref{fig:perceptual:staircase-procedure}.

Across the sequences, we present $5$ different reference colors, as visualized with black crosses in \Cref{fig:perceptual:thresholds}, each presented at $10^\circ$, $25^\circ$, and $35^\circ$ retinal eccentricities.
For each reference color, we adjust the color from four directions along the two equi-luminant cardinal axes in the \emph{DKL} color-space.
Briefly, \emph{DKL} is a perceptually uniform color space that is conducive to color vision research \cite{Krauskopf:1992:DiscriminationAdaptation,Hansen:2008:ColorNatural,Hansen:2009:ColorVsEccentricity}.
Please refer to \Cref{sec:method:color} for a detailed overview of the \emph{DKL} color space and why it is used in our perceptual studies.

The entire study took approximately 1.5 hours for each participant and they were encouraged to take breaks in between sequences. At the beginning of each user study, the participants completed $1$ sequence to familiarize with the procedure and equipment.

\subsection{Results}
In total, 8,123 trials were obtained from our participants (5 participants each with 60 sequences consisting of $\approx$21 trials each on average).
We record the color values at each reversal in \emph{DKL} coordinates, and average the last $3$ reversals (out of $6$ total) to determine the final discrimination threshold for each participant.
The average thresholds across all participants are visualized in \Cref{fig:perceptual:thresholds} in red, along with the $75\%$ confidence interval error bars.
As we approach the reference color from four directions in \emph{DKL} space, we obtain four different thresholds for each color at each eccentricity.
The lines connecting the four thresholds do not represent the shape of the overall threshold, and is only served as a visual guide to group each set of thresholds together. 
To avoid visual clutter, we plot discrimination thresholds at $10^\circ$ eccentricity for each reference color and $10^\circ$, $25^\circ$, and $35^\circ$ eccentricity thresholds for one reference color.
Refer to \Cref{sec:supp:pilot-individual} for all the measured threshold values separated by each participant.

\begin{figure*}
    \centering
    \subfloat[equipment setup]{
      \includegraphics[width=0.27\linewidth,trim=0 0 0 -2cm,valign=t]{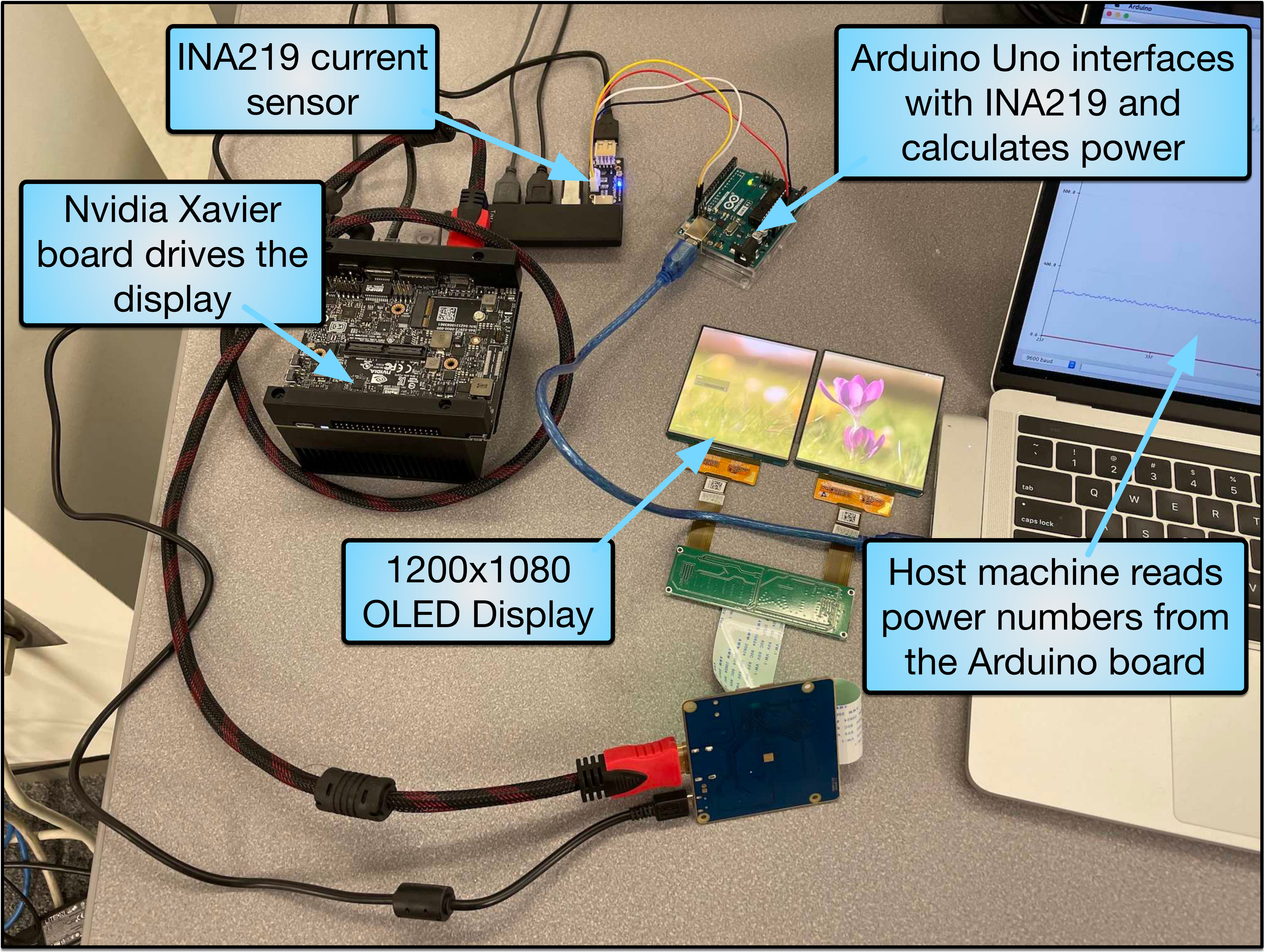}
      \vphantom{\includegraphics[width=0.346\linewidth,valign=t]{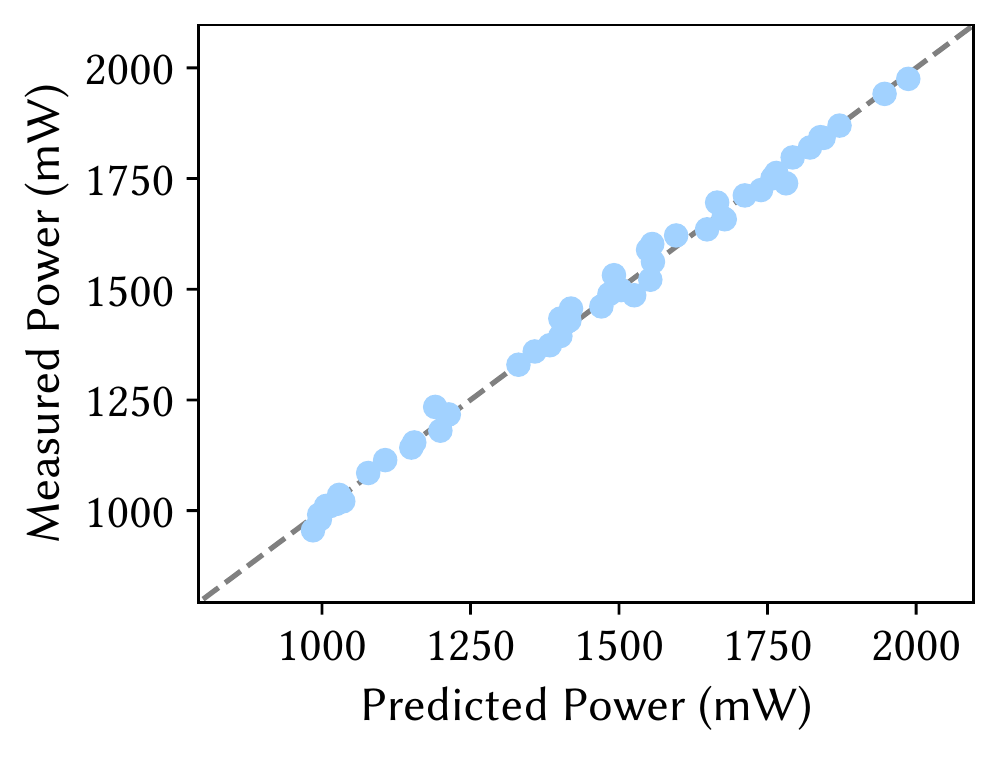}}
      \label{fig:energy:equipment-setup}
    }%
    \subfloat[measurement output]{
      \includegraphics[width=0.346\linewidth,valign=t]{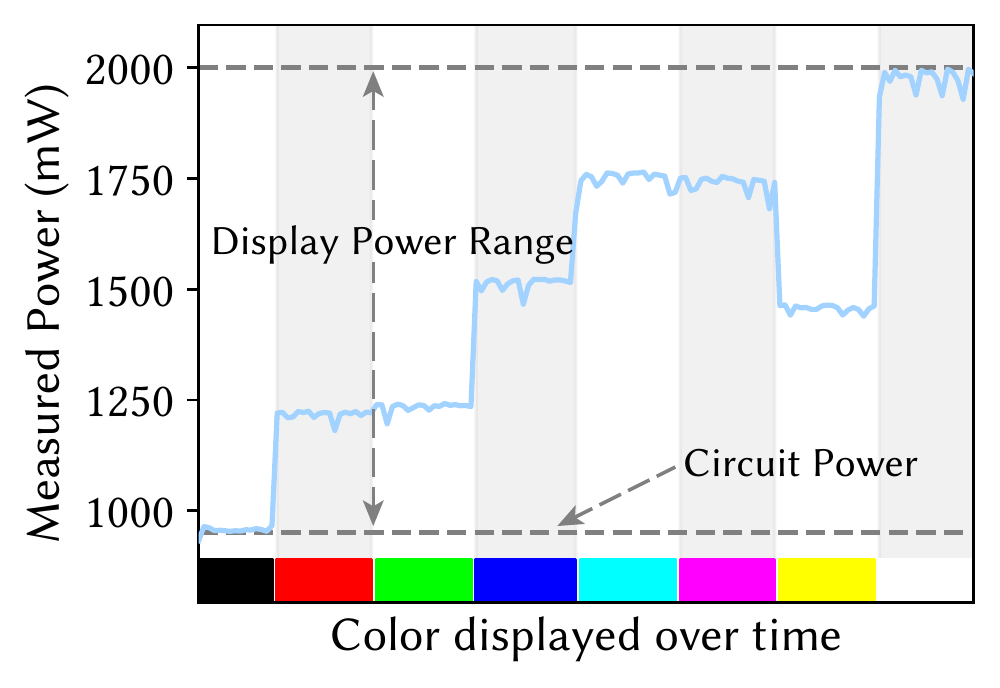}
      \vphantom{\includegraphics[width=0.346\linewidth,valign=t]{figures/model-energy/energy_model_fit.pdf}}
      \label{fig:energy:measurement-output}
    }%
    \subfloat[model accuracy]{
      \includegraphics[width=0.346\linewidth,valign=t]{figures/model-energy/energy_model_fit.pdf}
      \label{fig:energy:model-regression}
    }%
    \Caption{
        Display power measurement and modeling.
    }{
      \subref{fig:energy:equipment-setup}
      The display is connected to an Nvidia Xavier board, which provides power and frame data.
      The display power is intercepted by a Texas Instruments INA219 current sensor.
      The current and voltage readings are transmitted to an Arduino Uno board, which calculates the power.
      A host laptop obtains the power consumption from Arduino Uno through the serial port.
      \subref{fig:energy:measurement-output}
      The voltage and current readings are multiplied at each timestep to compute the power reading time-series.
      To measure the power for displaying different colors, we sample the \emph{sRGB} color space; each color is displayed for a period of 5 seconds. Here we show the power trace when cycling through the eight vertices of the \emph{sRGB} color cube.
      \subref{fig:energy:model-regression}
      We regress a linear power model by randomly sampling 52 colors in the \emph{sRGB} color space, and plot the measured power ($x$-axis) vs. the estimated power ($y$-axis).
      The mean relative error of the regression is $0.996\%$, indicating an accurate power model.
      The dashed line indicates the line of perfect agreement (i.e. $y=x$).
    }
    \Description{Display power measurement and modeling}
    \label{fig:energy}
\end{figure*}

\subsection{Discussions}
\label{sec:perceptual-study:discussion}
We chose the \emph{DKL} color-space as our color sampling space primarily because prior work suggests that the \emph{DKL} color space yields a perceptually uniform sampling space and, for that reason, is used extensively in perceptual studies \cite{Krauskopf:1992:DiscriminationAdaptation,Hansen:2008:ColorNatural,Hansen:2009:ColorVsEccentricity}.
For our work, we only sampled colors on a single equi-luminant plane.
In \emph{DKL} space, this corresponds to keeping the third dimension of the color space constant.
First, we observed unequal thresholds with different reference colors even if they were displayed at the same eccentricity.
However, they all appeared an ellipse shape, as also evidenced by prior literature \cite{Krauskopf:1992:DiscriminationAdaptation}.
That motivates us to develop our computational perceptual model considering the reference color as one of the inputs.

Prior work which utilizes the \emph{DKL} color-space suggests that discriminative thresholds measured with respect to a specific adaption luminance can be extended to arbitrary adaptation luminances due to the linearity of the cone-opponent process \cite{larimer1974opponent,larimer1975opponent}.
We use these results in this research and only conducted discriminative threshold measurements at a single adaptation luminance of $71.5$~cd$/\text{m}^2$ as mentioned above.

In the scope of our work, we did not study how spatial frequencies of stimuli affect discriminative thresholds.
Our experimental data provides the thresholds for a stimulus with a dominant frequency equal to $0.2$ cpd corresponding to the stimulus size used throughout the experiment.

Unsurprisingly, our data shows a decrease of ability to discriminate chromatic discrepancies as the retinal eccentricity increases.
The trend agrees with past experiments \cite{Hansen:2009:ColorVsEccentricity}, and is intuitive given the higher density of retinal receptors in the fovea \cite{Song:2011:ConeDensity}.
\Cref{fig:perceptual:thresholds} shows that the fall-off of discriminative sensitivity is very sharp, and the region of sub-threshold chromaticities at $35^\circ$ can take up as much as a third of the observable hues.
Some participants noted that at high eccentricities, all four disks appeared to be different, even though three of the disks were colored identically.
As such, the amount of noisy thresholds at high eccentricities attribute to the larger uncertainty for the overall threshold measurements as shown in \Cref{fig:perceptual:thresholds}.
Further investigations into this surprising phenomenon is an interesting future work.

We also observe inter-subject variation in the measured thresholds, as shown in \Cref{sec:supp:pilot-individual}.
While this could be due to a number of reasons (e.g., observer metamerism~\cite{xie2020observer}, prereceptoral filtering~\cite{norren1974spectral}, calibration, experimental setup, etc.), further study is required to understand the reason for these differences. {Nevertheless, for developing a computational model, we use the most conservative thresholds across participants, instead of an average fit. This assures
generalization to a larger population considering individual variances (see \Cref{sec:impl:preprocess})}.

Lastly, it is notably critical that those thresholds only hold for discriminative tasks. 
Using the observed thresholds, we performed a preliminary validation with a \emph{sequential detection task} and two-alternative-forced choice (2AFC). 
In this study, the same group of participants was instructed to observe pairs of stimuli and identify whether they appear identical.
Some of the trials consist of one non-altered image, with the other containing peripheral color altering within the identified thresholds. We observed that a majority of users can successfully identify the altered condition, suggesting the distinct perceptual thresholds between discrimination and detection tasks.
Nevertheless, during active vision tasks where an observer is instructed to freely observe natural visual content, their sensitivity may significantly reduce \cite{Cohen:2020:TLC}. 
We hypothesize that the color sensitivity during active vision is also lower than during discriminative tasks. We investigate and validate the hypothesis in more detail in \Cref{sec:result:perceptual}.

\section{Pilot Study: Measuring Display Power with Varied Colors}
\label{sec:display-energy-study}

To measure the power consumption characteristics of VR displays and how it varies depending on the images displayed on them, we conduct a hardware study, and later use the collected data to derive a model for predicting the power consumption of a display given the image displayed on it.

\subsection{Setup}
For our power study, we use the Wisecoco H381DLN01.0 OLED \cite{wisecocoampled}. The display module has two identical displays, each with a resolution of 1080$\times$1200, matching the aspect ratio of HTC Vive Pro Eyes, which is what we use for perceptual studies. 

We do not use the native display modules in Vive Pro Eye HMD and Oculus Quest 2 for power studies, because their displays are physically tightly integrated into the headsets; thus, the display power cannot be easily isolated from the rest of the system. In the case of the Oculus Quest 2, the headset is powered by a battery that is tightly integrated into the headset, which prevents us from using methods used in studying smartphone display power, where the battery is unplugged and replaced with an external power supply that has internal power sensing capabilities~\cite{Dash2021,halpern2016mobile,dong2009power}.

\Cref{fig:energy:equipment-setup} shows the experimental setup to measure display power. We intercept the display power supply with a SwitchDoc PowerCentral board, which has an on-board INA219 module (with a 0.1$\Omega$ shunt resistor) to measure the current. The INA219 module is connected to an Arduino board through the I2C interface. We develop a driver that runs on the Arduino board to get the display current and voltage, from which we can calculate the power.

The driver running on the Arduino board configures the INA219 sensor to output a new power measurement every $\sim 68$ ms; each power reading is internally averaged over 128 samples, resulting in an effective power sampling rate of $\sim 1,882$ Hz.

\subsection{Measurement and Discussion}
As a preliminary test, we measure the power consumption of the eight vertices of the \emph{sRGB} color cube. For each color, we set all the display pixels to that color, display it for five seconds, and calculate the average power.
\Cref{fig:energy:measurement-output} shows the measured power trace. It is clear that the display power consumption is sensitive to the color.

We make two observations from \Cref{fig:energy:measurement-output}. First, even when the display is showing black pixels, i.e., when the LEDs are not emitting light, there is a non-trivial amount of \emph{static} power consumption. The power beyond the static portion is consumed by the LEDs, which we dub the dynamic display power.
This static power is consumed by the peripheral circuitry that drives the LEDs, such as the per-pixel transistors and capacitor as well as the addressing logic~\cite{huang2020mini}. The contribution of the static power is about 50\% in display white and is about 80\% when displaying red and green.

The trend of semiconductor technology is that the circuit power is decreasing over time with better fabrication technologies~\cite{bohr200730}, but the LED power is much harder to reduce because the display must sustain certain luminance levels to meet brightness requirements, which arguably do not change dramatically over time. Our work aims to reduces the (color-sensitive) dynamic power of the display, which will become more important as the static power reduces in the future.

Second, the dynamic power consumption of red and green colors are roughly half that of blue. This is because displaying the \emph{sRGB} blue on our display requires contributions from both the blue and red sub-pixels (due to the primaries used by this display) as confirmed by examining the microscopic images of the display (\Cref{fig:energy:leds}). As a result, if we expect to see any energy wins, we anticipate that green-, and/or, red-shifting images can decrease the power consumption of the image. 
We will leverage the measured data to obtain a computational power-vs-color model in \Cref{sec:method:energy}.

\begin{table}
    \Caption{Our microscopic photos of the display under different colors.}{
    We image the display under \emph{sRGB} red, green, blue, and white colors using a Carson MicroFlip mircoscope with a magnification of 120x. One can observe that the display red and green primaries roughly match the red and green primaries in the \emph{sRGB} color space, but the \emph{sRGB} blue requires contributions from both the blue and red sub-pixels from the display.
    }
\begin{tabular}{ c c c c }
    red & green & blue & white\\
    \hline
    \\
    \begin{minipage}{0.20 \linewidth}
        \includegraphics[width=\linewidth]{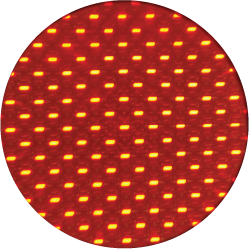}
    \end{minipage}
    &
    \begin{minipage}{0.20 \linewidth}
        \includegraphics[width=\linewidth]{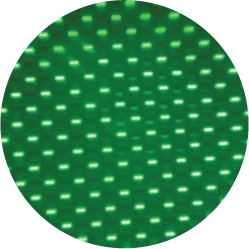}
    \end{minipage}
    &
    \begin{minipage}{0.20 \linewidth}
        \includegraphics[width=\linewidth]{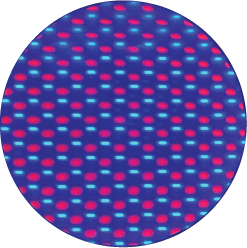}
    \end{minipage}
    &
    \begin{minipage}{0.20 \linewidth}
        \includegraphics[width=\linewidth]{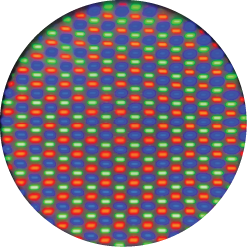}
    \end{minipage}
    \\
    \\
    \hline
\end{tabular}
    \label{fig:energy:leds}
\end{table}

\begin{figure*}
    \centering
    \subfloat[model prediction at $10^\circ$ \& $25^\circ$ in \emph{DKL} space]{
      \includegraphics[width=0.31\linewidth,valign=t]{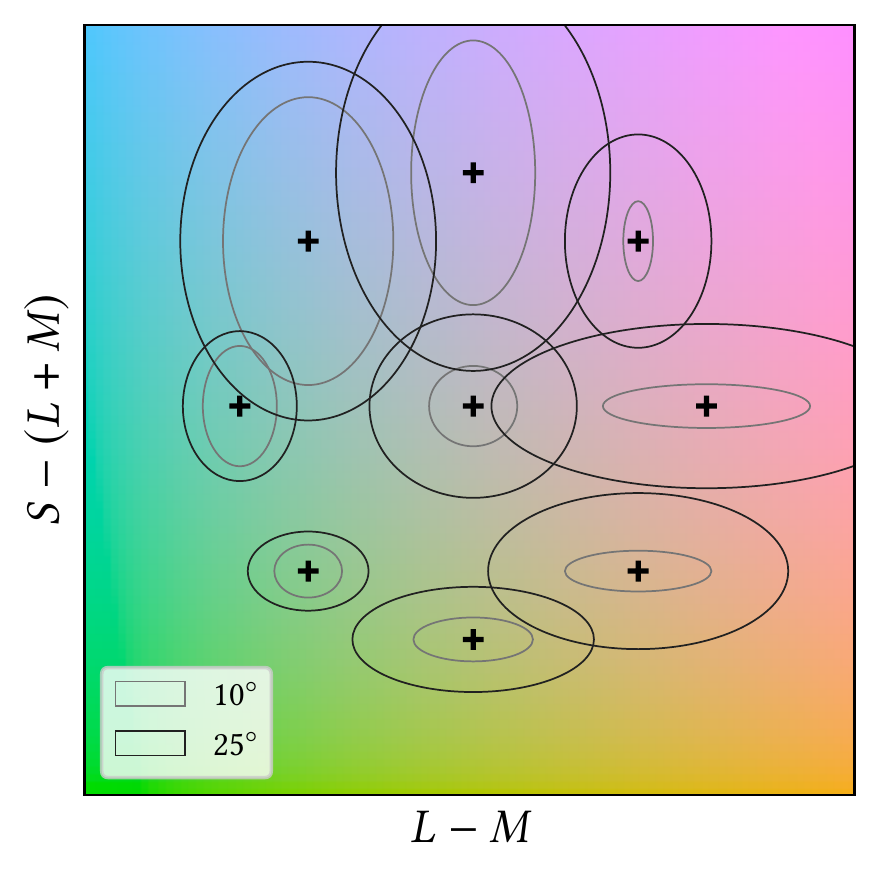}
      \label{fig:model:dkl-predictions}
    }%
    \subfloat[model prediction at $25^\circ$ in linear \emph{sRGB} space]{
      \includegraphics[width=0.32\linewidth,valign=t]{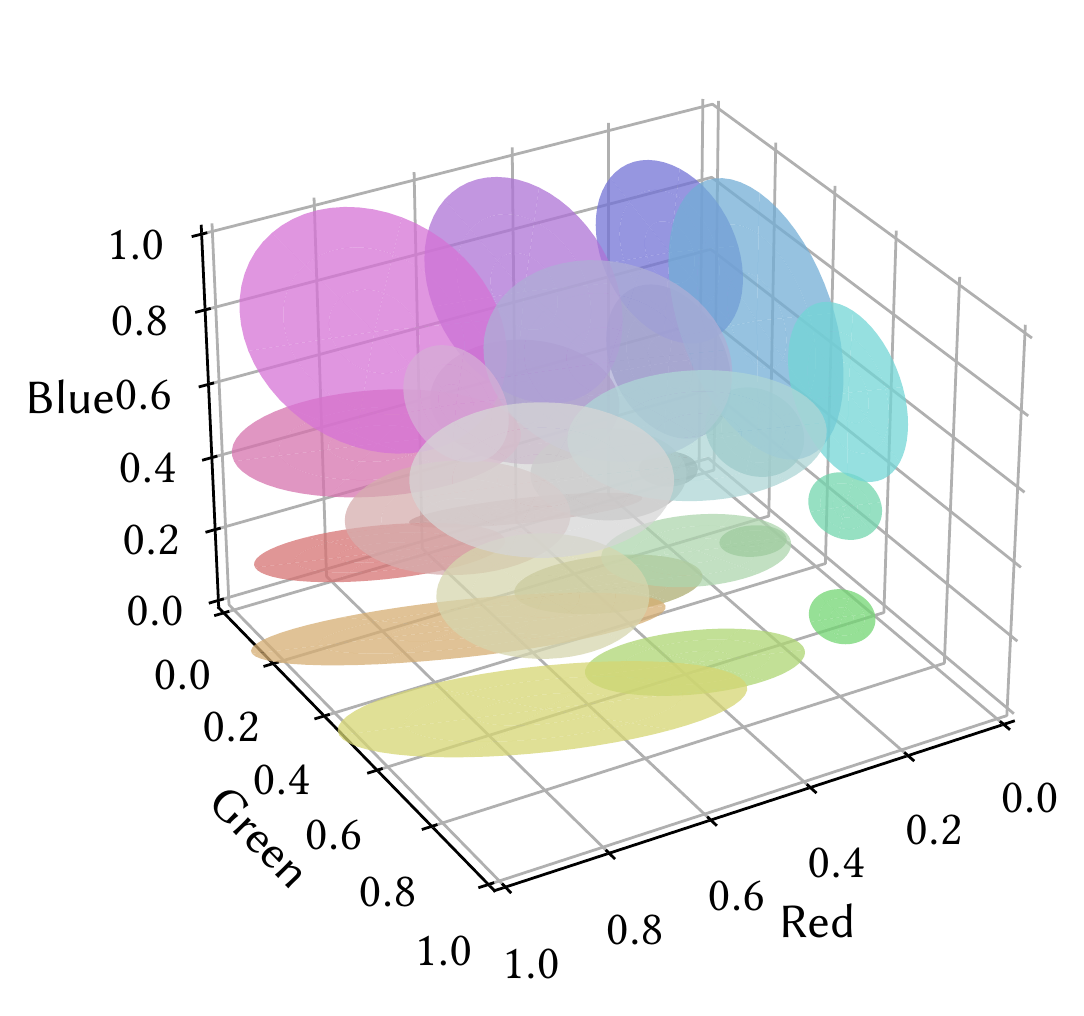}
      \vphantom{\includegraphics[width=0.31\linewidth,valign=t]{figures/color-model/color_thresholds_dkl.pdf}}
      \label{fig:model:srgb-predictions}
    }%
    \subfloat[vector field for power optimization]{
      \includegraphics[width=0.32\linewidth,valign=t]{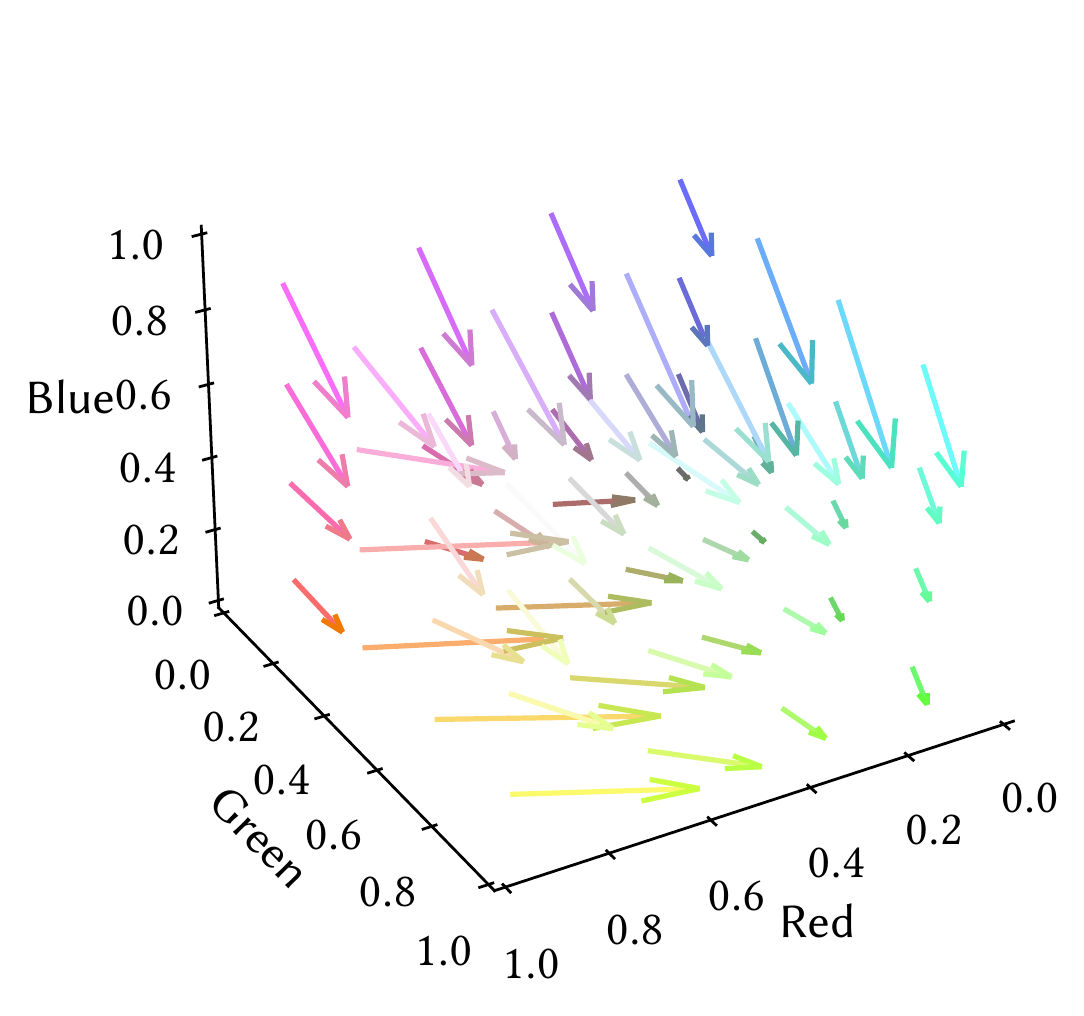}
      \vphantom{\includegraphics[width=0.31\linewidth,valign=t]{figures/color-model/color_thresholds_dkl.pdf}}
      \label{fig:model:vector-field}
    }%
    \Caption{
        Color perception model and power-aware chromaticity optimization.
    }{
        \subref{fig:model:dkl-predictions} illustrates our color discrimination model's ellipse thresholds at nine coordinates.
        They are evaluated on an equi-luminant plane in the \emph{DKL} color space, with eccentricities $10^\circ$ (light gray curve) and $25^\circ$ (dark black curve).
        All colors within an ellipse are perceptually indistinguishable from the center (black-cross) color.
        Similarly, these ellipse thresholds can also be sampled in linear \emph{sRGB} space.
        \subref{fig:model:srgb-predictions} shows ellipses of samples across a $3\times3$ grid sampled within the \emph{sRGB} cube.
        Each ellipse in this illustration is shaded with the color at its center.
        Since equi-luminant planes are parallel in linear \emph{sRGB}, all the ellipses appear on parallel planes.
        \subref{fig:model:vector-field} visualizes the model-guided chromaticity shifting (at $25^{\circ}$ eccentricity) to minimize display power consumption. We use our perceptual constraints in combination with our power cost function to shift the chromaticities of various sample colors inside the \emph{sRGB} color cube (illustrated in \Cref{fig:optimize}).
        The original, and power-optimized colors correspond to the tail and head of the vectors, respectively.
    }
    \Description{Perceptual Model Predictions}
    \label{fig:perceptualModel}
\end{figure*}

\section{Model: Perceptually Guided Power Optimization}
\label{sec:method}
Using the results of our perceptual user study, and hardware power measurements, we develop a display power optimization model under the constraint that the change in the images observed by human subjects is not perceptible. In \Cref{sec:method:color}, we first derive a computational model of human color discrimination (\Cref{fig:perceptualModel}) using the data obtained from \Cref{sec:perceptual-study}. In \Cref{sec:method:energy}, we build a linear power consumption model regressed from the physical measurement data in \Cref{sec:display-energy-study}. Finally, in \Cref{sec:method:optimization}, we integrate the two models above (as a constrained convex optimization) toward a closed-form display color modulation function. It aims to minimize the display's power consumption while ensuring the modulation within the human discriminative thresholds.

\subsection{Perceptual Model for Color Discrimination}
\label{sec:method:color}
The study of human color vision suggests that the physiological mechanisms which govern our discriminative ability of colors rely on comparing the different cone receptor cell activations in our retinas \cite{DeValois:1966:LGNanalysis}.
Specifically, the two primary comparison mechanisms are the difference between (1) the $L$ vs $M$ cone and (2) the $L+M$ vs $S$ cone activations \cite{Derrington:1984:DKL}.
These observations are the foundation of the \emph{cone opponent} theory of color discrimination and are widely accepted for measuring human color discriminative thresholds \cite{Conway:2018:ColorVisionTour}.

In this research, we leverage these results from the vision science community and develop a computational framework that quantifies the discriminative threshold of any given color at different retinal eccentricities.
That is, given a colored stimulus at some retinal eccentricity, we determine the extent to which we can modulate its color while maintaining its perceptual color appearance to an observer.

The set of colors which are indistinguishable from some \emph{test} color by human observers are modeled as ellipse shaped regions defined over equi-luminant color-spaces \cite{Krauskopf:1992:DiscriminationAdaptation, Hansen:2008:ColorNatural, Hansen:2009:ColorVsEccentricity}.
Notably, the MacAdam ellipses~\shortcite{Macadam:1942:Ellipses} are the first to model discriminative thresholds as such.
Additionally, Krauskopf and Karl~et~al.~\shortcite{Krauskopf:1992:DiscriminationAdaptation} show that the sizes of these ellipses are best described in the \emph{DKL} color-space \cite{Derrington:1984:DKL}. We first introduce the \emph{DKL} space, followed by how we model the discriminative threshold in the \emph{DKL} space.

\paragraph{DKL color space.}
\emph{DKL} space defines colors in two steps. First, it quantifies a color using the cone-opponent mechanism. Therefore, a color is first defined over a basis that computes the cone-opponent neural activations.
Namely, a color with \emph{LMS} coordinates $(\testCoordinate_L, \testCoordinate_M, \testCoordinate_S)$ is first converted to the basis:
\begin{align}
\begin{aligned}
\testCoordinate_{L-M} &= \testCoordinate_L - \testCoordinate_M\\
\testCoordinate_{S-(L+M)} &= \testCoordinate_S - (\testCoordinate_L + \testCoordinate_M)\\
\testCoordinate_{L+M} &= \testCoordinate_L + \testCoordinate_M,
\label{eq:method:perceptual:linear-dkl}
\end{aligned}
\end{align}

Second, instead of directly using the cone-opponent basis, the \emph{DKL} space models colors in terms of \emph{color contrast}; that is, colors are defined relative to a reference (a.k.a., the \emph{adaptation}) color. This is different from the more absolute measure of color used in \emph{sRGB}, \emph{XYZ}, and \emph{LMS} color spaces, where each color is defined based on its own characteristics.
Specifically, given a \emph{test} color, $\testVec$, and an \emph{adaptation} color, $\adaptationVec$, we can compute the color-contrast of the test color with respect to the adaptation color as:
\begin{align}
\colorContrastCoordinate(\testCoordinate_i; \adaptationCoordinate_i) = \frac{\testCoordinate_i - \adaptationCoordinate_i}{\adaptationCoordinate_i}
\text{, for } i \in \{1, 2, 3\}.
\label{eq:method:perceptual:color-contrast}
\end{align}
Depending on the specific basis used to represent these colors, the indices $\{1, 2, 3\}$ could be $\{X, Y, Z\}$ for the \emph{XYZ} basis or $\{L, M, S\}$ for the \emph{LMS} basis. In our case, the indices $\{1, 2, 3\}$ are $\{L-M, S-(L+M), L+M\}$, since we have already defined colors in the cone-opponent space, as shown in \Cref{eq:method:perceptual:linear-dkl}.

It is worth noting \Cref{eq:method:perceptual:linear-dkl} can be seen as producing an intermediate color space that is a linear transformation away from the conventional \emph{LMS} space.
We will henceforth refer to the intermediate color space given by \Cref{eq:method:perceptual:linear-dkl} as \emph{i(ntermediate)-DKL}.

We use the \emph{LMS} color space as defined by Smith and Pokorny~\shortcite{Smith:1975:LMS}, which is what the original \emph{DKL} space is based on~\cite{Derrington:1984:DKL}.
The particular \emph{LMS} cone fundamentals are so defined that the coordinate $\testCoordinate_{L+M}$ of a color is strictly equal to the luminance of the color, i.e., the $Y$ coordinate in the \emph{XYZ} space.

\paragraph{Modeling ellipse level sets.}

In our model, we represent the set of all equi-lumiant colors which cannot be discriminated from a test color, $\testVec \in \text{\emph{i-DKL}}$, relative an adaptation color, $\adaptationVec \in \text{\emph{i-DKL}}$, using an ellipse-shaped region centered around the color contrast of the test color.
The boundary of this ellipse region corresponds to the discriminative threshold of $\colorContrastCoordinate(\testCoordinate_i, \adaptationCoordinate_i)$, for $i \in \{L-M, S-(L+M)\}$.
The set of color coordinates which represent this threshold, $\colorVec \in \text{\emph{i-DKL}}$, fulfill the system of equations:
\begin{align}
\begin{cases}
\colorCoordinate_{L+M} &= \adaptationCoordinate_{L+M}\\
\ellipse(\colorVec; \testVec, \adaptationVec, \normalizedEllipseVec) &= 0.
\end{cases}
\label{eq:method:perceptual:constraint-eqn}
\end{align}
The first constraint ensures that all the color coordinates on the threshold are equi-luminant to the adaptation color.
The second constraint ensures that all $\colorVec$ are on the edge of the ellipse region with major and minor semi-axes equal to $\normalizedEllipseVec = (\normalizedEllipseCoordinate_{L-M}, \normalizedEllipseCoordinate_{S-(L+M)}) \in \real^2$.
Formally, the function $\ellipse(\cdot)$ is defined as
\begin{align}
\ellipse(\colorVec; \testVec, \adaptationVec, \normalizedEllipseVec)
= \sum_{i = \{L-M, S-(L+M)\}}
\left(
\frac{
    \colorContrastCoordinate(\colorCoordinate_i; \adaptationCoordinate_i) -
    \colorContrastCoordinate(\testCoordinate_i; \adaptationCoordinate_i)
}{
    \normalizedEllipseCoordinate_i
}
\right)^2
- 1,
\label{eq:method:perceptual:constraint-contrast}
\end{align}

\paragraph{Model Regression.}
\Cref{eq:method:perceptual:constraint-contrast} requires the knowledge of the ellipse-size parameters, $\normalizedEllipseCoordinate_i$.
Prior work has shown that $\normalizedEllipseCoordinate_i$ relates to the color-contrasts of various test colors, $\colorContrastCoordinate(\testCoordinate_i, \adaptationCoordinate_i)$, as well as the retinal eccentricity, $\eccentricity \in \real^+$, at which a colored stimulus is displayed \cite{Krauskopf:1992:DiscriminationAdaptation,Hansen:2009:ColorVsEccentricity}.
We leverage our user study data from \Cref{sec:perceptual-study} to learn the relationship
\begin{align}
\neuralNetwork : \ (\colorContrastVec, \eccentricity) \mapsto \normalizedEllipseVec
\end{align}
where $\colorContrastVec \in \real^2$ are the $L-M$ and $S-(L+M)$ coordinates of the test color in \emph{DKL} space computed using \cref{eq:method:perceptual:linear-dkl,eq:method:perceptual:color-contrast}.
Specifically, we use our data to optimize a shallow neural network, which estimates the discrimination thresholds, using least-squares regression:
\begin{align}
\perceptualModel = \argmin_{\neuralNetwork} \lVert\neuralNetwork(\colorContrastVec, \eccentricity) - \normalizedEllipseVec\rVert_2^2.
\end{align}
The $R^2$ value of the regression is $0.58$ (adjusted $R^2$ value of $0.51$), indicated an acceptable regression accuracy.
Note that our raw data from \Cref{sec:perceptual-study} is intentionally pre-processed as described in detail in \Cref{sec:impl:preprocess}.
Briefly, we aim to cover more \emph{conservative} thresholds that are generalizable to broad users instead of an ``average fit''.

\paragraph{Neural Network Architecture.}
We chose the Radial Basis Function Neural Network (RBFNN) with a sigmoid output layer to ensure local smoothness, as well as a positive, localized output range.
Mathematically, the network is summarized as
\begin{align}
\neuralNetwork(\colorContrastVec, \eccentricity)
= \ellipseMax \odot S\left(
    \sum_{j=1}^{\rbfNodes}
        \rbfWeight_j
        \gaussianBasis \left(
            \left\lVert
            \genfrac[]{0pt}{0}{\colorContrastVec}{\eccentricity} - \rbfCentre_j
            \right\rVert_2,
            \rbfStd_j
        \right)
        + \rbfBias_j
    \right),
\label{eq:method:perceptual:rbfnn}
\end{align}
where $\odot$ is the term-wise multiplication operator.
The RBFNN takes the input, and computes the weights of the effect each of the $\rbfNodes$ nodes of the latent representation have on the input.
It does so by applying a Gaussian Radial Basis function, $\gaussianBasis$, centered at $\rbfCentre_j$ with std of $\rbfStd_j$, for each node, $j$.
The weights of each node is scaled by a scaling constant $\rbfWeight_j$, incremented by the linear bias $\rbfBias_j$, summed up, and passed to the sigmoid function $S$ and mutliplied by a scaling factor $\ellipseMax$ to produce the final prediction.
The trainable parameters of this network are the centres, $\rbfCentre_j$, sizes, $\rbfStd_j$ of the radial bases, as well as the final scaling factors $\rbfWeight_j$, and linear biases $\rbfBias_j$.
$\ellipseMax$ is a normalization constant and chosen to be the maximum possible value of contrasts within the capability of the display used in our work, and hence does not change.
For our work we keep the number of nodes $N = 5$ low to maintain smoothness of the outputs.
Please refer to our source code for more details on the model specifications.

\paragraph{Ellipse re-parameterization.}
Since the adaptation color, $\adaptationVec$, is the same for all variables in \cref{eq:method:perceptual:constraint-contrast}, we simplify the function by re-parameterization as $\ellipseCoordinate_i = \normalizedEllipseCoordinate_i \adaptationCoordinate_i$ for $i \in \{L-M, S-(L+M)\}$:
\begin{align}
\ellipse(\colorVec; \testVec, \ellipseVec)
= \sum_{i = \{L-M, S-(L+M)\}}
\left(
\frac{
    \colorCoordinate_i - \testCoordinate_i
}{
    \ellipseCoordinate_i
}
\right)^2
- 1.
\label{eq:method:perceptual:constraint}
\end{align}

While the original formulation in \cref{eq:method:perceptual:constraint-contrast} relates the ellipse to variables in \emph{DKL} space, and are ultimately the variables used to regress the model, as we'll see in \Cref{sec:method:energy}, it's helpful to reformulate the model with respect to the differences of color in \emph{i-DKL}: $\colorVec - \testVec$, as well as the new parameter, $\ellipseVec$, which clearly represents the size of the ellipse in \emph{i-DKL} space, are both measures defined within \emph{i-DKL}.
In summary, \cref{eq:method:perceptual:constraint} converts \cref{eq:method:perceptual:constraint-contrast} from a \emph{DKL} space parameterization to a \emph{i-DKL} parameterization. The obtained model is visualized in \Cref{fig:model:dkl-predictions,fig:model:srgb-predictions}.

\subsection{Power Model for Display Illumination}
\label{sec:method:energy}

In this section we derive a computation model that correlates an OLED's power consumption with the pixel color.
The display power is modeled as the sum of the LED power, which consists of the powers of its three sub-pixels, and the power of the peripheral circuitry (e.g., the thin-film transistors)~\cite{huang2020mini}.
It is known that the power of an OLED sub-pixel is roughly proportional to its current, which is proportional to the numerical value of the corresponding channel \cite{tsujimura2017oled}.
Thus, given the \emph{RGB} value of the three sub-pixels, $\colorVec_{disp} \in \text{\emph{disp-RGB}}$ (i.e., the pixel value in the display native color space),
its total power consumption is
\begin{align}
\power
= \left(\sum_{i=\{1, 2, 3\}} \powerCoordinate_i \colorCoordinate_i\right) + \powerCirc
= \powerVec_{disp}^T \colorVec_{disp} + \powerCirc,
\label{eq:method:power:total-power}
\end{align}
where $\powerVec_{disp} \in \real^3$ is the vector of unit powers of each sub-pixel, and $\powerCirc \in \real$ is the static power consumption (consumed by the peripheral circuits) when all the pixels are black, i.e., the LEDs do not emit light and, thus, do not consume power.

In most computer graphics applications, it is impractical to use the display's native color space because it varies depending on the manufacturer specifications, and could be unknown.
Color-spaces that are commonly used, such as (linear) \emph{sRGB}, can transform to a display's native color-space via some linear transformation, $\colorSpaceTransform \in \real^{3\times3}$.
Without loss of generality, using this transformation, we can rewrite \cref{eq:method:power:total-power} in terms of the (linear) \emph{sRGB} pixels as
\begin{align}
\begin{aligned}
\power(\colorVec_{srgb})
&= \powerVec_{disp}^T \colorSpaceTransform_{srgb2disp} \colorVec_{srgb} + \powerCirc\\
&= \powerVec_{srgb}^T \colorVec_{srgb} + \powerCirc,
\label{eq:method:power:srgb-power}
\end{aligned}
\end{align}
where $\colorSpaceTransform_{srgb2disp}$ is the transformation matrix from (linear) \emph{sRGB}'s color-space to the display's, and $\colorVec_{srgb} \in \text{{sRGB}}$ denotes the pixel color in linear \emph{sRGB} space.
For convenience, we define $\powerVec_{srgb}^T = \powerVec_{disp}^T \colorSpaceTransform_{srgb2disp}$, which intuitively denotes the power consumption of the three display sub-pixels under unit \emph{sRGB} simuli.

$\powerVec_{srgb}$ depends on the specification of a particular display.
In our work, we study an OLED display module from Wisecoco that has two 1080×1200 displays, as described in \Cref{sec:display-energy-study}. 
Critically, our methodology is not unique to the specific display at study and, thus, can be extended to build power models for any other three-primary display.

\paragraph{Power model regression.}
To build an analytical power model, we must find the parameter, $\powerVec_{srgb}$.
We do so by physically measuring the power consumption of 52 randomly sampled colors in the \emph{sRGB} space, including the eight colors that correspond to the eight vertices of the \emph{sRGB} color cube, as described in \Cref{sec:display-energy-study}, and solving an over-determined linear system,
\begin{align}
\power^{(color)} = \powerVec_{srgb}^T \colorVec_{srgb}^{(color)} + \powerCirc,
\end{align}
where $color$ is the 52 sampled colors, via the classic linear least squares method.
\Cref{fig:energy:model-regression} shows the measured power of these sampled colors ($y$-axis) and the regressed model outputs ($x$-axis). The mean relative error of the regression is $0.996\%$, indicating an accurate model.

\subsection{Optimizing Display Energy Consumption under Perceptual Constraints}
\label{sec:method:optimization}

Finally, using \cref{eq:method:perceptual:constraint-eqn} and \cref{eq:method:power:srgb-power} we can minimize the power consumption function of a display, $\power(\colorVec)$, while constrained within the perceptual limits set by $\ellipse(\colorVec)$.
Qualitatively, we notice that the power function is a linear function of the input, $\colorVec$, so the minimizing power will be on the surface of the discriminative threshold ellipse (as opposed to its interior).
Notice that in this optimization problem, it is more convenient to use the definition of the ellipses in \emph{i-DKL} space, instead of the \emph{DKL} space definition (cf. \cref{eq:method:perceptual:constraint} and \cref{eq:method:perceptual:constraint-contrast}) because the \emph{i-DKL} space is only a linear transformation away from (linear) \emph{sRGB}, therefore making its energy computations a lot simpler.

Formally, we define the optimization process as:
\begin{align}
\begin{aligned}
&\colorVec_{idkl}^* = \argmin_{\colorVec_{idkl}} \power(\colorSpaceTransform_{idkl2srgb}\colorVec_{idkl})\\
&\text{subject to: } \ellipse(\colorVec_{idkl}; \testVec_{idkl}, \ellipseVec = \normalizedEllipseVec \odot \adaptationVec_{idkl}) = 0,
\label{eq:method:power:opt}
\end{aligned}
\end{align}
where the original color of the pixel is $\testVec$ and, the adaptation color of the display is $\adaptationVec$.
In our work we choose $\adaptationVec$ to be equal to a color with a chromaticity equal to the CIE D65 Standard Illuminant (i.e., the reference white in the \emph{sRGB} color space) and a luminance equal to the luminance of the test pixel $\testVec$.
While the choice of adaptation color is an interesting question to explore, it is beyond the scope of this work and is left as future work.

\begin{figure}
    \centering%
    \includegraphics[width=0.8\linewidth]{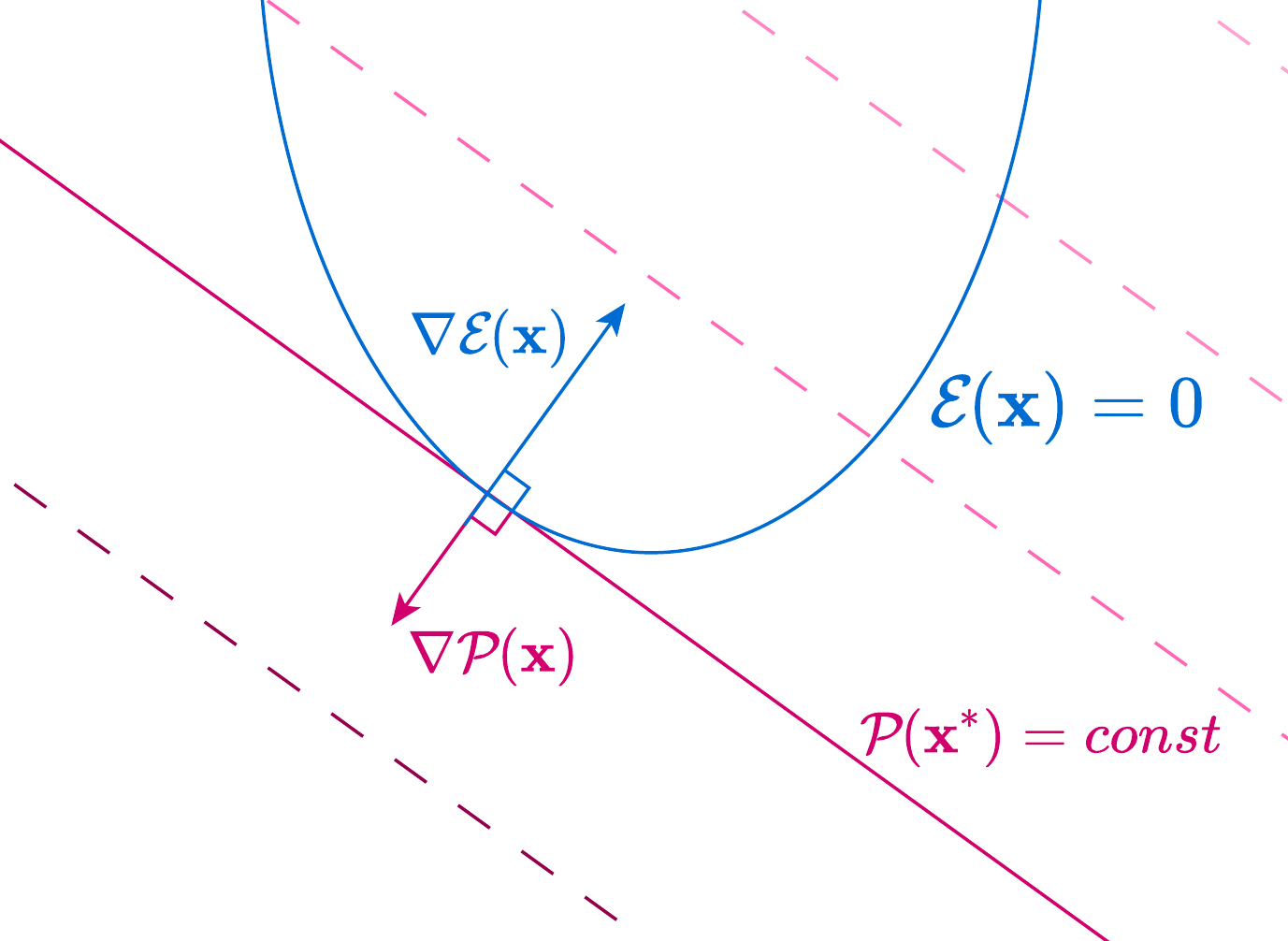}%
    \Caption{%
        Illustration of deriving the closed-form and optimal chromaticity.%
    }{%
        We aim to search the color $\colorVec^*$ which (1) minimizes our first-order power function, $\power(\colorVec)$; (2) is under the constraint $\ellipse(\colorVec)=0$ when the gradients of both functions are co-linear.
        The straight lines represent the set of all colors consuming the same amount of power.
        Since the color-spaces \emph{sRGB} and \emph{i-DKL} are related via a linear transformation, this solution works in either color-space.
    }%
    \label{fig:optimize}%
\end{figure}

Due to the convexity of both the cost and constraint functions, we can apply the method of Lagrange multipliers to find the output color, $\colorVec_{srgb}^*$, which minimizes the total power consumption in closed form:
\begin{align}
\begin{aligned}
\colorVec_{srgb}^* = \colorSpaceTransform_{idkl2srgb}
\begin{bmatrix}
\frac{\powerCoordinate_1 \ellipseCoordinate_1^2}{\sqrt{\powerCoordinate_1^2\ellipseCoordinate_1^2 + \powerCoordinate_2^2\ellipseCoordinate_2^2}} \\
\frac{\powerCoordinate_2 \ellipseCoordinate_2^2}{\sqrt{\powerCoordinate_1^2\ellipseCoordinate_1^2 + \powerCoordinate_2^2\ellipseCoordinate_2^2}} \\
\testCoordinate_3
\end{bmatrix},
\end{aligned}
\end{align}
where $\{1, 2, 3\}$ correspond to $\{L-M, S-(L+M), L+M\}$.
\Cref{fig:optimize} visually illustrates how this optimal color is found using the derivatives of $\ellipse$, and $\power$.
Please refer to \Cref{appn:optimal-derivation} for the derivation of the above result.

\section{Implementation}
\subsection{Perception Study Data Pre-processing}
\label{sec:impl:preprocess}

We take two steps to pre-process the perception study data. Both steps are meant to keep the model's threshold estimation conservative, which is necessary for two reasons. First, there are natural variances across participants (\Cref{sec:perceptual-study:discussion}) and, thus, a conservative estimation allows our model to generalize to large populations. Second, our model is built to modulate the displayed colors to preserve the visual fidelity in active viewing, which we hypothesize to have a lower threshold than that in discriminative tasks.

First, we use the smallest thresholds across participants, instead of an average fit.
Second, we observed small asymmetries in the collected thresholds, and we confirmed that this is also the case in Krauskopf and Karl et al.~\shortcite{Krauskopf:1992:DiscriminationAdaptation}.
We made the engineering decision to keep our model's thresholds more conservative;
thus each threshold is chosen to be the narrower one from the two thresholds approached from opposing sides along a \emph{DKL} axis.
That is, given a threshold approached from the positive $L-M$ side, $\normalizedEllipseCoordinate_{L-M}^+$\revision{}{,} and one from the negative $L-M$ side, $\normalizedEllipseCoordinate_{L-M}^-$, the discrimination threshold we pick for model regression is:
\begin{align}
    \normalizedEllipseCoordinate_{L-M} = \min ( \normalizedEllipseCoordinate_{L-M}^+, \normalizedEllipseCoordinate_{L-M}^- ),
\end{align}
and \revision{vice-versa}{similarly} for the $S - (L + M)$ axis.

\subsection{Eccentricity Extrapolation}
\label{sec:impl:eccentricity}
In our perceptual model regression, we restricted the range of valid input eccentricities to be between $10^\circ$ and $35^\circ$ because we had only measured discriminative thresholds within this range of eccentricities.
We avoided color-shifting content at eccentricities $<10^\circ$ due to the low power-saving payoffs for foveal and para-foveal regions.
Meanwhile, eccentricities $>35^\circ$ were clamped down to $35^\circ$ as a conservative estimate.

\subsection{Shader}
\label{sec:impl:shader}
We implement a post-processing image-space shader in the Unity ShaderLab language to compute per-pixel power-minimizing color.
\Cref{fig:shader} outlines the pseudocode of our shader.
We tested our shader on the HTC Vive Pro Eye (relevant specs shown below) powered by an NVIDIA RTX3090 GPU, and observed that processing each frame takes less than $11$ ms, which ensures no loss of frames in the displays.
\begin{center}
\captionof{table}{Relevant specifications for the HTC Vive Pro Eye}
\begin{tabular}{c|c}
    Feature & Specification\\
    \hline
    Display Resolution & $1440\times1600$ pixels per eye\\
    Display Refresh-rate & $90$ Hz\\
    Peak Luminance & $143$ cd$/\text{m}^2$\\
    Eye-tracker Accuracy & $0.5^\circ - 1.1^\circ$\\
    Eye-tracker Frequency & $120$ Hz.
\end{tabular}
\end{center}
\begin{figure}
\begin{tcolorbox}[
    enhanced,
    attach boxed title to top left={xshift=6mm,yshift=-3mm},
    colback=moonstoneblue!20,
    colframe=moonstoneblue,
    colbacktitle=moonstoneblue,
    title=Power-Minimizing Color Computation,
    fonttitle=\bfseries\color{black},
    boxed title style={size=small,colframe=moonstoneblue},
    sharp corners,
]
\medskip
\begin{algorithmic}[1]
    \Function{Fragment\_Shader}{}
        \State $\textsc{pos}_\textsc{pixel} = $ \Call{GetPixelPos}{}
        \State $\textsc{pos}_\textsc{gaze} = $ \Call{GetGazePos}{}
        \State $\testVec_{\textsc{srgb}} = $ \Call{SampleTexture}{MainTex, $\textsc{pos}_\textsc{pixel}$}
        \State $e = $ \Call{GetEccentricity}{$\textsc{pos}_\textsc{gaze}, \textsc{pos}_\textsc{pixel}$}
        \If{$e < \textsc{ecc}_\textsc{min}$}
            \State \textbf{return} $\testVec_\textsc{srgb}$
        \ElsIf{$e > \textsc{ecc}_\textsc{max}$}
            \State $e = \textsc{ecc}_\textsc{max}$
        \EndIf
        \State $\testVec_{\textsc{idkl}} = \colorSpaceTransform_{\textsc{srgb2idkl}} \testVec_{\textsc{srgb}}$
        \State $\textsc{lum}=\testVec_{\textsc{idkl}}$.z
        \State {\color{blue}\(\triangleright\) Adaptation color}
        \State $\adaptationVec_{\textsc{idkl}}=\colorSpaceTransform_\textsc{srgb2idkl} [\textsc{lum},\textsc{lum},\textsc{lum}]^T$
        \State {\color{blue}\(\triangleright\) $\odot$: element-wise multiply}
        \State $\colorContrastVec=(\testVec_\textsc{idkl} - \adaptationVec_\textsc{idkl}) \odot \frac{1}{\adaptationVec_\textsc{idkl}}$
        \State $\textsc{input}=\genfrac[]{0pt}{0}{\colorContrastVec}{\eccentricity}$ {\color{blue} \Comment{Model input}}
        \State \Call{Init}{} $\textsc{rbf}[5]$, $\textsc{linear}[2]$ {\color{blue} \Comment{Model output}}
        \State {\color{blue}\(\triangleright\) Lines 14-22: RBFNN from equation \eqref{eq:method:perceptual:rbfnn}}
        \For{$i$ in [0, 5)}
            \State $\textsc{rbf}_i\gets
                     \gaussianBasis \left(
                                        \left\lVert
                                        \textsc{input} - \rbfCentre_i
                                        \right\rVert_2,
                                        \rbfStd_i
                                    \right)$
        \EndFor
        \For{$i$ in [0, 2)} {\color{blue} \Comment{Linear layer}}
            \State $\textsc{linear}_i\gets \rbfWeight_i \odot \textsc{rbf} + \rbfBias_i$
        \EndFor
        \State $\normalizedEllipseVec \gets \ellipseMax \odot \Call{Sigmoid}{\textsc{Linear}}$
        \State $\ellipseVec = \normalizedEllipseVec \odot \adaptationVec_\textsc{idkl}$
        \State {\color{blue}\(\triangleright\) Compute power-optimal color}
        \State $\powerVec = $ \Call{GetPowerModelCoeffs}{}
        \State $\colorVec^*_\textsc{idkl} = \testVec_\textsc{idkl} + \ellipseVec^2 \odot \powerVec \frac{1}{\sqrt{\sum_i\ellipseCoordinate_i^2 \powerCoordinate_i^2}}$
        \State \textbf{return} $\colorSpaceTransform_\textsc{idkl2srgb} \colorVec^*_\textsc{idkl}$
    \EndFunction
\end{algorithmic}
\end{tcolorbox}
\Caption{%
Shader implementation pseudocode.
}{%
The shader routine optimizes an input color into the optimized color as described by our method from \Cref{sec:method}.
Refer to our source code for the ShaderLab implementation.
}
\label{fig:shader}
\end{figure}

\section{Evaluation}
\label{sec:result}
In this section we evaluate the performance and applicability of our model.
In \Cref{sec:result:perceptual}, we first conduct a psychophysical experiment to assess the perceptual fidelity between our method and an alternative luminance-based power reduction approach. The experiment design follows previous literature \cite{Cohen:2020:TLC}.
Then, in \Cref{sec:result:imagenet}, we measure the model's generic benefits in broad applications by further analyzing the display power saving ratio with a large scale natural image
dataset,
ImageNet.

\subsection{Psychophysical Study for Perceptual Quality}
\label{sec:result:perceptual}
\begin{figure*}
    \centering
    \subfloat[original]{
      \includegraphics[width=0.22\linewidth]{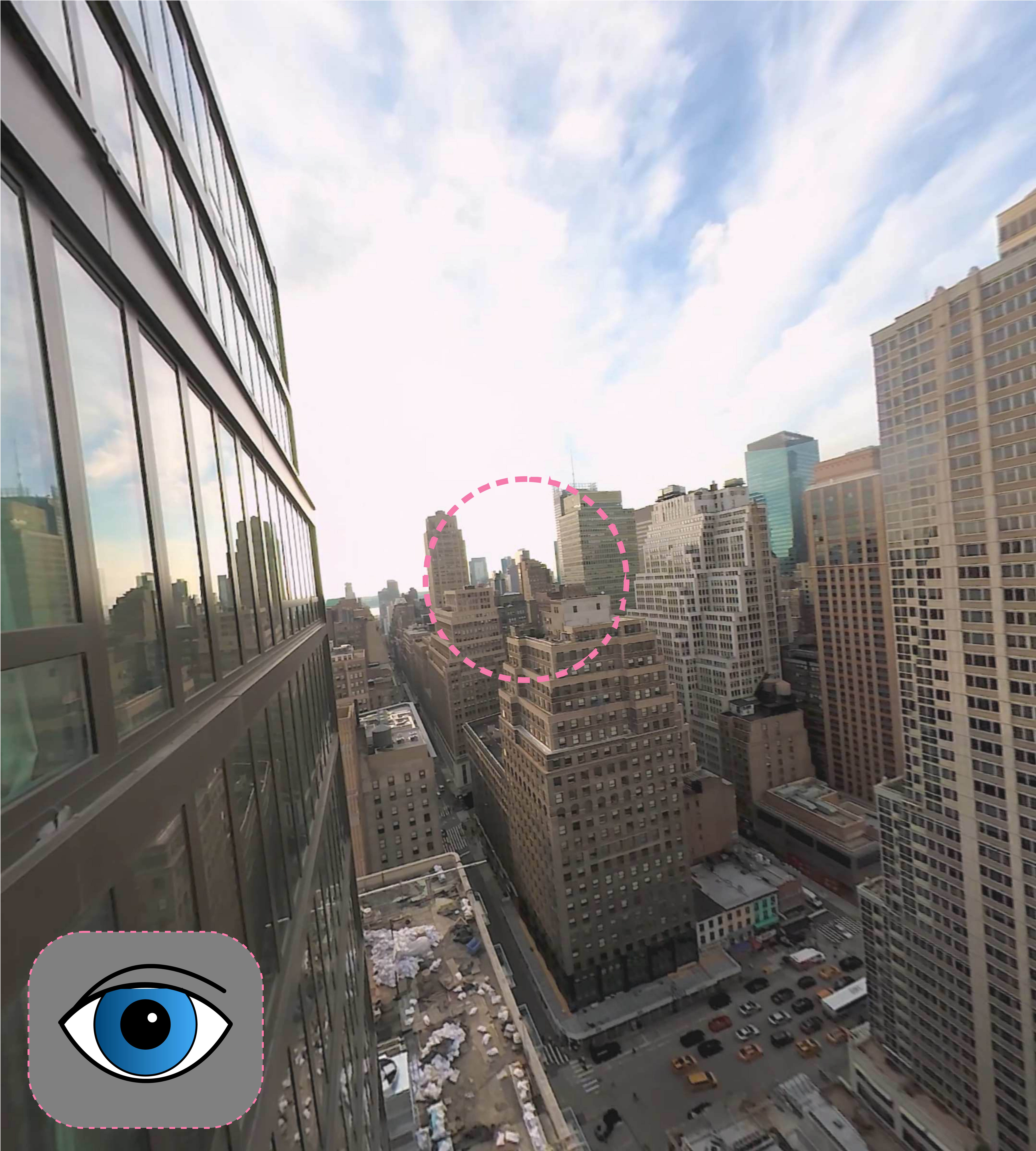}
      \label{fig:eval:original}
    }%
    \subfloat[our method, $\conditionOurs$]{
      \includegraphics[width=0.22\linewidth]{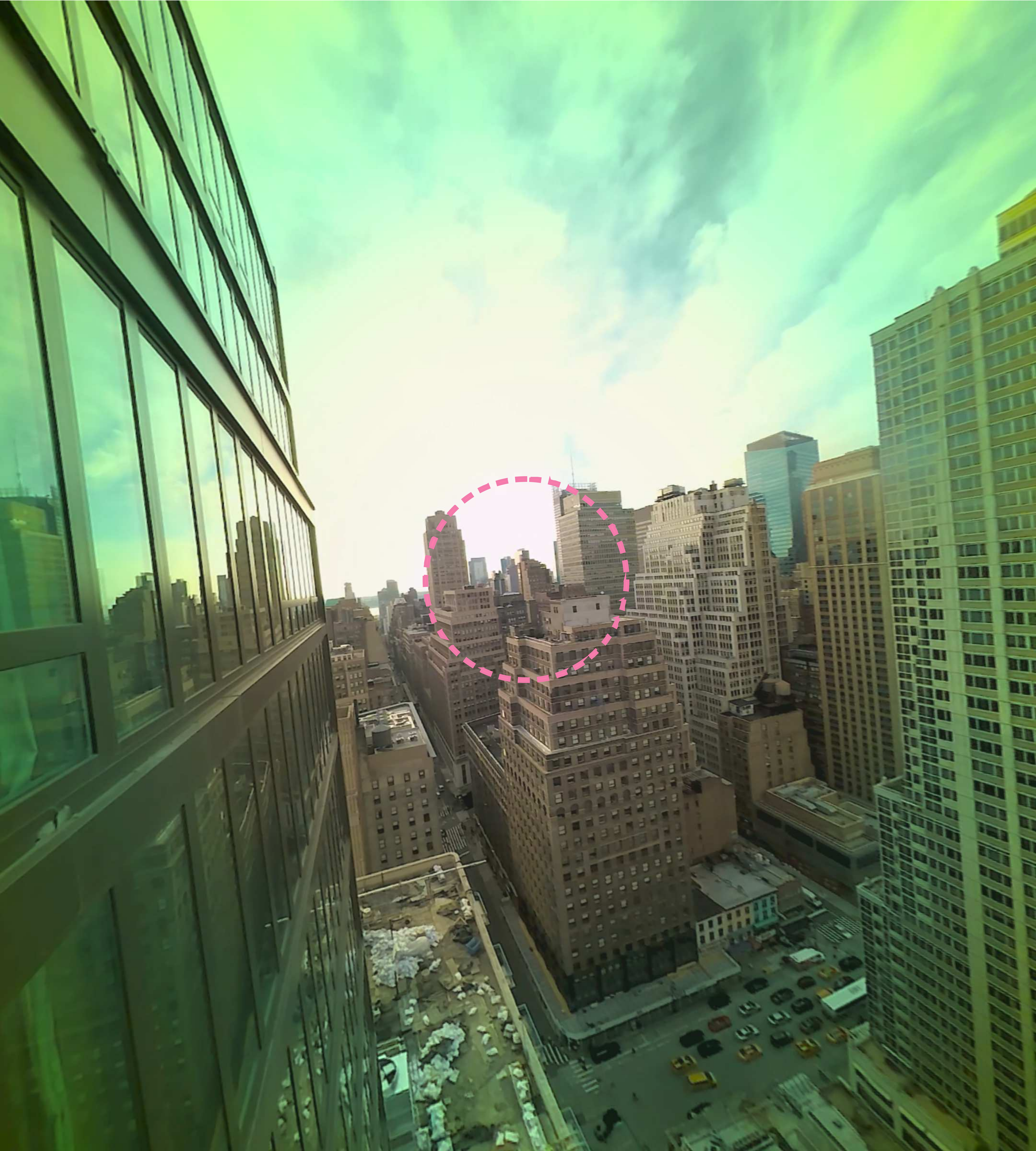}
      \label{fig:eval:ours}
    }%
    \subfloat[luminance-based, $\conditionUniform$]{
      \includegraphics[width=0.22\linewidth]{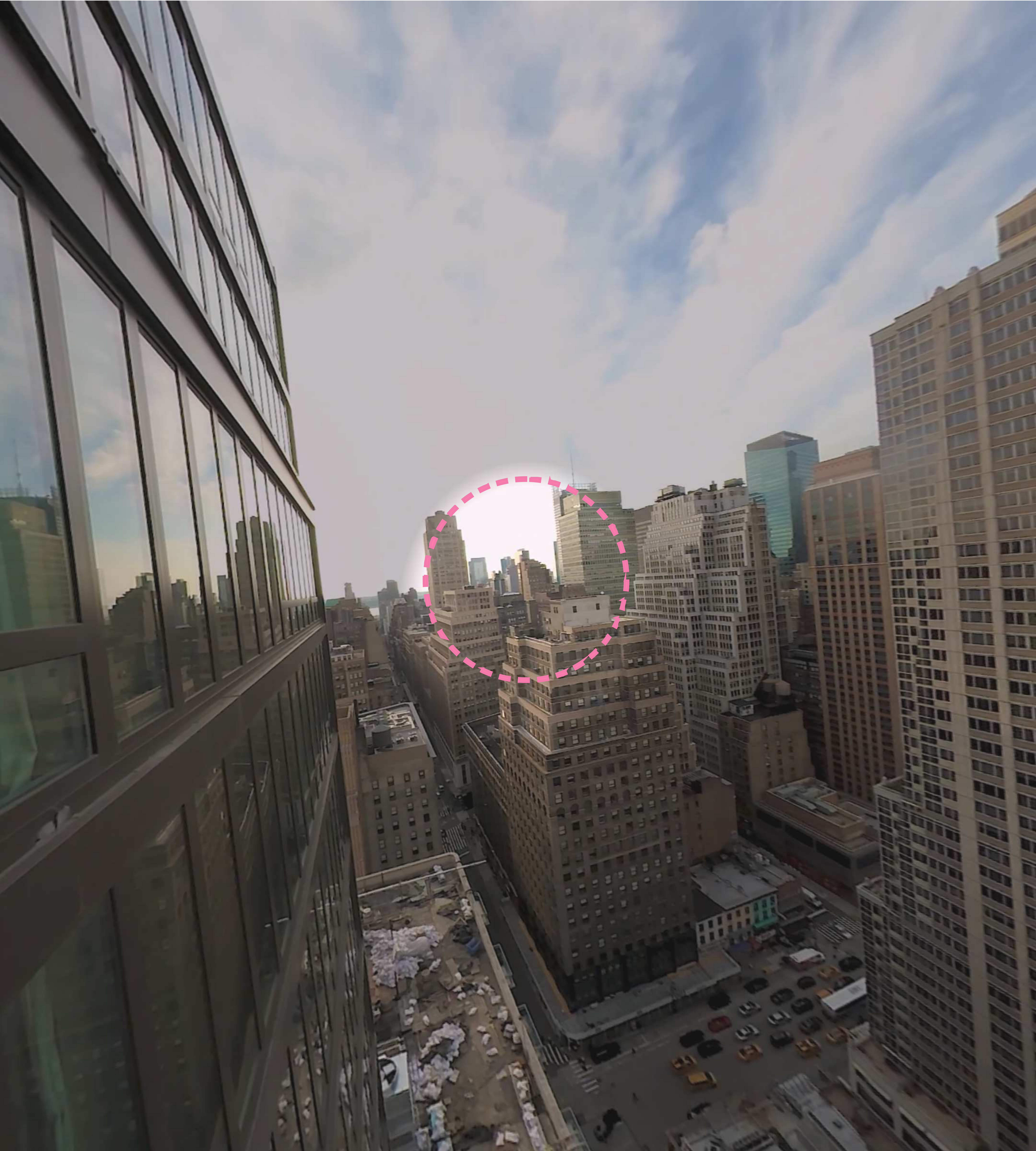}
      \label{fig:eval:uniform}
    }%
    \subfloat[temporal onset]{
      \includegraphics[width=0.33\linewidth]{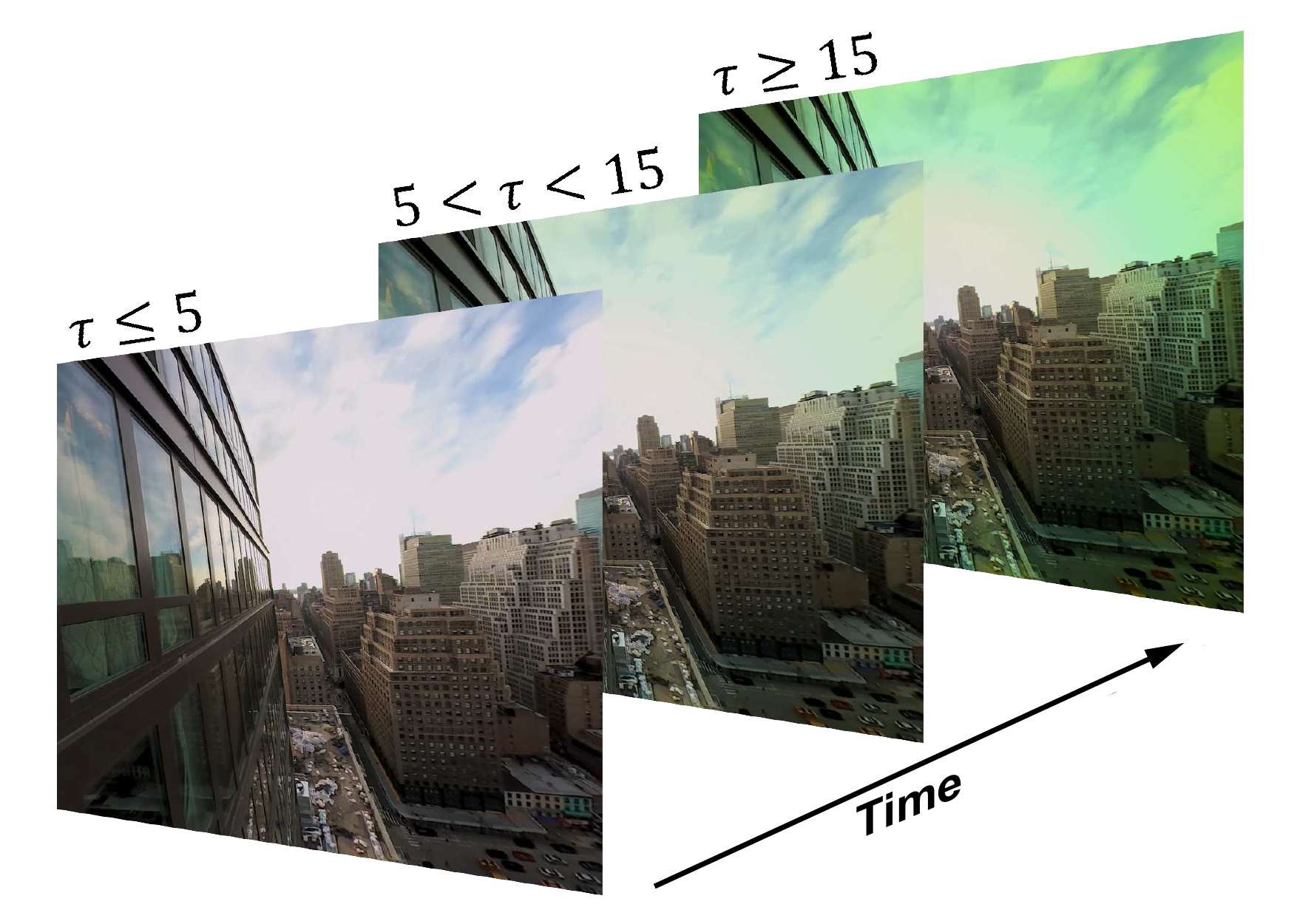}
      \label{fig:eval:temporal}
    }%

     \subfloat[perceptual quality]{
      \includegraphics[width=0.5\linewidth]{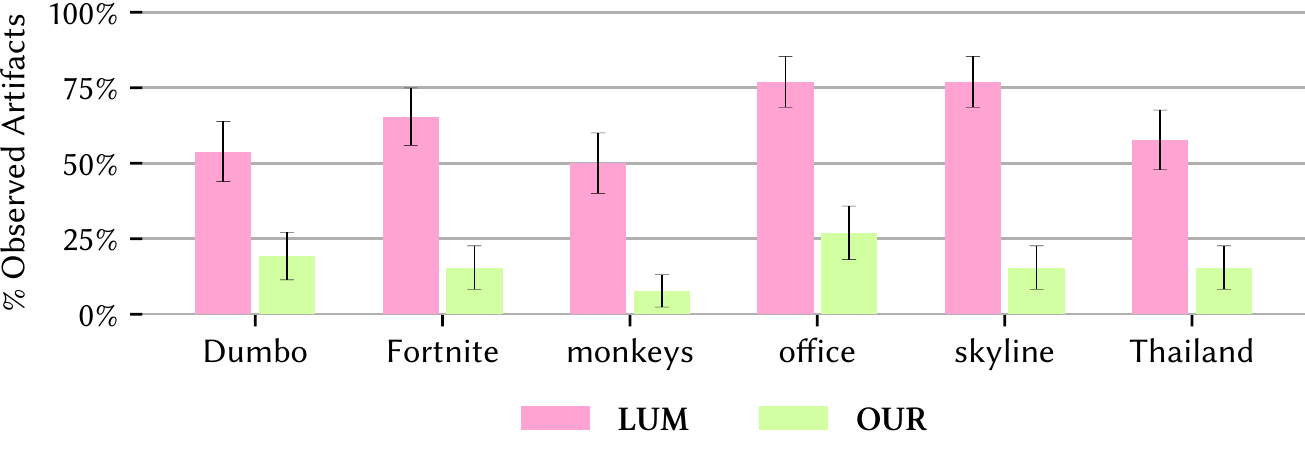}
      \label{fig:eval:bar-perceptual}
    }%
    \subfloat[display power saving]{
      \includegraphics[width=0.5\linewidth]{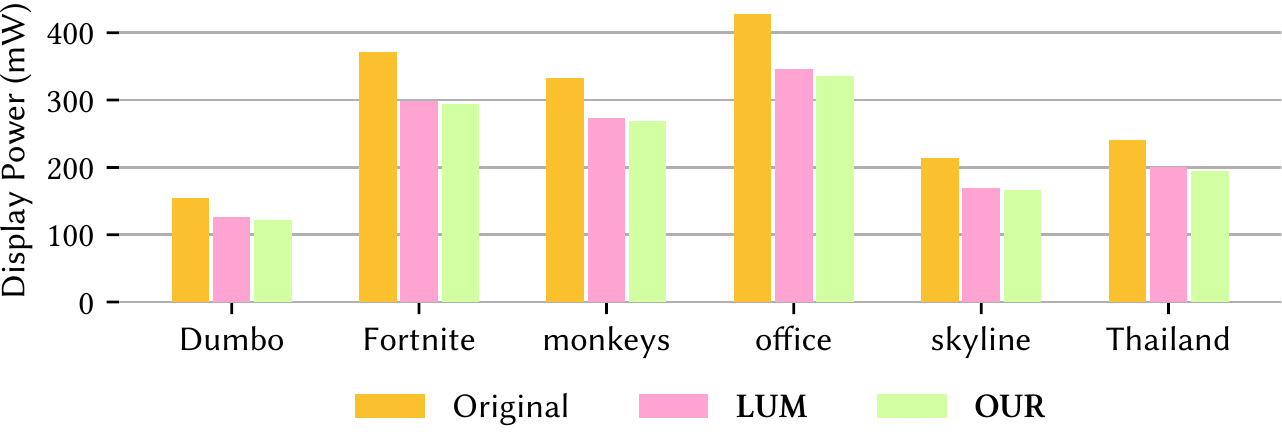}
      \label{fig:eval:bar-power}
    }%
    \Caption{
        User study stimuli and results.
    }{
    \subref{fig:eval:ours} and \subref{fig:eval:uniform} show the results of applying the two gaze-contingent shading conditions $\conditionOurs$ and $\conditionUniform$ to an example frame \subref{fig:eval:original} in the video sequence stimuli, with the dashed circles indicating the user's gaze.
    \subref{fig:eval:temporal} We increase the intensity of the gaze-contingent filter in a temporal fashion, while keeping the foveal region unchanged. From 0-5 seconds, the video is unfiltered. Between 5-15 seconds we gradually insert one of the two conditions, with the remaining time between 15-20 seconds the full filter is applied. Please refer to our supplementary video for a dynamic visualization of this temporal process.
    \subref{fig:eval:bar-perceptual}
    shows the percentage of trials where users identified ``artifacts''. 
    Across the 6 scenes, $\conditionOurs$ exhibits significantly lower values than $\conditionUniform$, evidencing our method's benefit in preserving perceptual fidelity. 
    The error bars indicate standard error across users.
    \subref{fig:eval:bar-power} 
    Using the hardware setup in \Cref{sec:display-energy-study}, we physically measure the scene-dependent display power (measured power excluding the $\sim 1000$mW peripheral circuit power) consumption of each 20-second video clip among three conditions: the original, $\conditionUniform$, and $\conditionOurs$. $\conditionOurs$ achieves similar power saving capability to $\conditionUniform$.
    Unmodified image credits to Humaneyes Technologies.
    }
    \Description{Evaluation scenes}
    \label{fig:result:eval}
\end{figure*}

Motivated by the experiment of Cohen et~al.~\shortcite{Cohen:2020:TLC}, we conduct a psychophysical user study to measure participant-experienced fidelity deterioration, as well as the corresponding power-saving level during active and real-world viewing. 
``Active and real-world'' is notably a condition where participants may freely rotate their head/eyes and naturally investigate an immersive scene. 

\paragraph{Setup and participants}
We recruited $13$ participants (ages 21-32, 3 female). 
None of the participants were aware of the research, the hypothesis, or the number of conditions.
All participants have normal or corrected-to-normal vision.
We used the same hardware setup as our preliminary user study in \Cref{sec:perceptual-study}. 
Before each experiment, we ran a five-point eye-tracking calibration for each participant.

\paragraph{Stimuli and conditions}
\begin{figure}
  \centering
  \includegraphics[width=\linewidth]{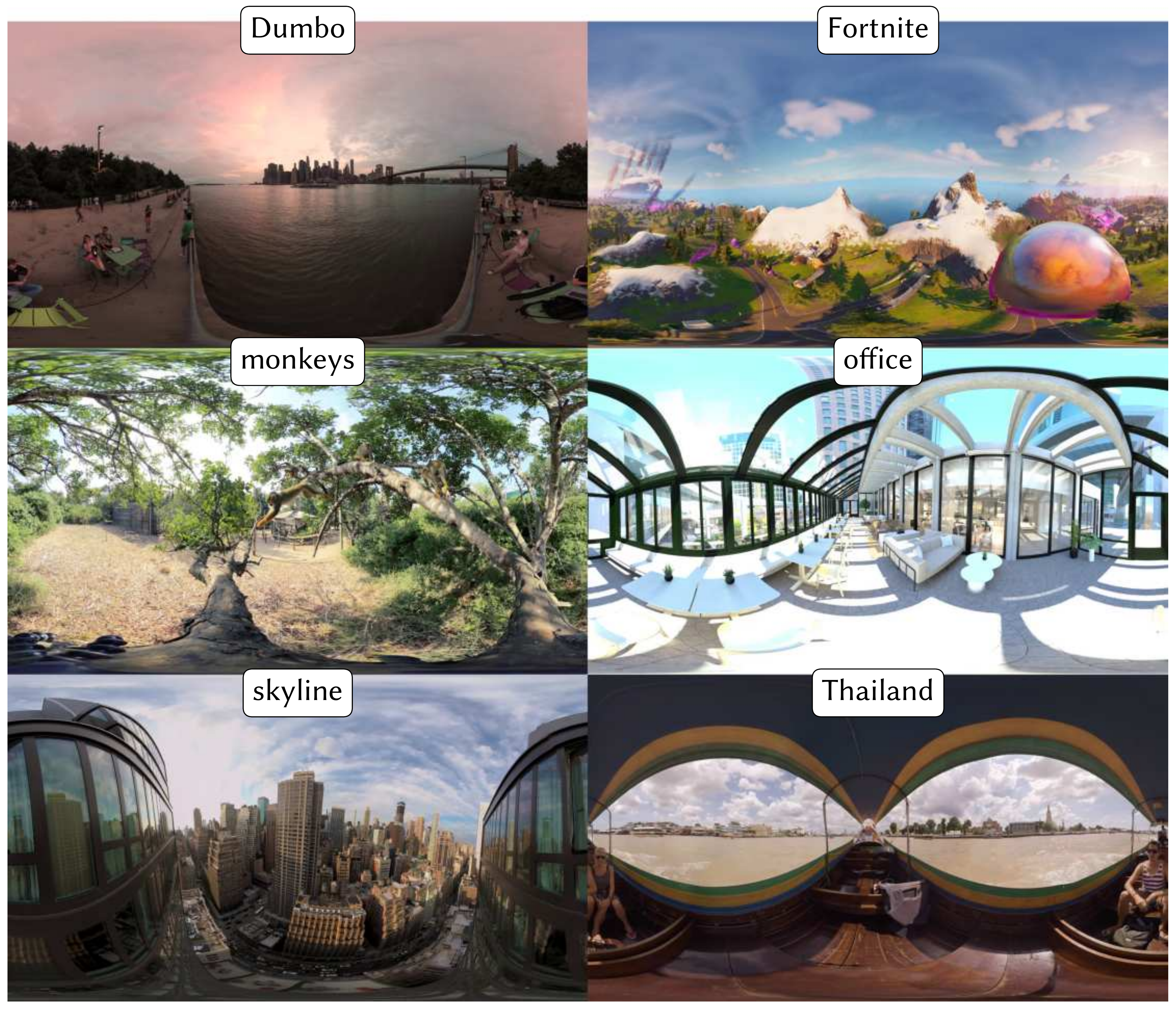}
  \Caption{
        Panoramic Video Scenes.
    }{
        Representative panoramic frames captured from 360 degree monoscopic video scenes used in the evaluation user study described in \Cref{sec:result:perceptual}. Images of "Dumbo", "monkeys", and "skyline" scenes by Humaneyes Technologies, rendering of historical Two Embarcadero Center in San Francisco, CA ("office") by Rene Rabbitt of Rabbitt Design, "Thailand" by VR Gorilla, and "Fortnite" by AmiramiX.
    }
  \label{fig:scenes}
\end{figure}

The stimuli were $6$ panoramic video sequences as shown in \Cref{fig:scenes}.
For broader coverage, the tested scenes contain natural/synthetic, static/dynamic, bright/dark, and indoor/outdoor content.

We studied the perceptual quality by applying two gaze-contingent and power-saving shading approaches to the scenes: a baseline luminance-modulated shader, $\conditionUniform$; and the shader with our chromaticity modulation method, $\conditionOurs$ (\Cref{sec:impl:shader}). 
Specifically, in $\conditionUniform$, we applied a constant scaling factor to all peripheral (eccentricity > $10^{\circ}$) pixels' colors. That is, $\conditionUniform$ can be understood as a gaze-contingent version of the ``power-saving mode'' on mobile devices.
The scaling factor was determined in such a way that the power saving (estimated using the power model) of $\conditionUniform$ is similar to that of $\conditionOurs$.
An example frame of the original stimulus, $\conditionOurs$, and $\conditionUniform$ are shown in \Cref{fig:result:eval}.

Similar to \cite{Cohen:2020:TLC}, we \emph{temporally} inserted one of the two shaders to the original stimulus during each trial. 
More formally, let $I_o$ be the original video and $I_p$ be the power optimized version.
Then starting at timestamp $\tau_o=5$s, we linearly interpolate (i.e. $lerp$) between $I_{o}$ and $I_{p}$ over a course of $10$ seconds.
At $\tau_p=15$s, the transition completes and the power optimized video is played henceforth.
That is,
\begin{align}
I(\tau) = lerp\left(I_o, I_p, \frac{\tau - \tau_o}{\tau_p - \tau_o}\right)
\end{align}
The process is illustrated in \Cref{fig:eval:temporal}. Note that the temporal insertion also implicitly compares the original frame with each of the two power-saved conditions.

\paragraph{Tasks}
Our experiment consisted of 24 trials {($6$ scenes $\times$ $2$ condition $\times$ $2$ repetitions)}, lasting approximately 15 minutes for each participant. 
Before the experiment started, we first displayed 2 trial runs to familiarize the participant with the setup. 
Afterwards, six 20-second video sequences (with representative frames displayed in \Cref{fig:scenes}) were shown to the participant in a counter-balanced randomized order.

During each trial, the participant was instructed to perform a scene-specific task, such as ``count the number of chairs'' to ensure they were actively viewing the scene. 
After each trial, the participant was instructed to answer both the scene-specific task and a two-alternative forced choice (2AFC, similar to \cite{Cohen:2020:TLC}) question ``did you notice any visual artifacts?''. 
Before beginning the experiment, we show the participants static frames from the ``skyline'' video as visual examples of ``artifact'' stimuli, including one original frame and the two corresponding conditions,  $\conditionOurs$, and $\conditionUniform$. 
The example images were displayed on a computer monitor (as opposed to the VR headset); thus, the participants' retinal image was significantly different from the stimuli shown during the study. This is to ensure that the participants are not biased when shown artifacts.

\paragraph{Metrics and results}
\label{sec:result:eval}
We use the percentage of trials where participants noticed artifacts as the metric of perceptual quality. Lower values indicate better quality, i.e., less noticeable visual modulation.
\Cref{fig:eval:bar-perceptual} plots the user-reported values of each scene and each condition.
As visualized in \Cref{fig:eval:bar-perceptual}, the average percentage of observed artifacts in $\conditionUniform$ is $63.5 \pm 9.4\%$ (STE) and in $\conditionOurs$ is $16.7 \pm 7.3\%$.
The lowest percentage of observed artifacts in scenes with $\conditionOurs$ applied occurred in the monkeys scene, a scene with large amounts of green, whereas the highest percentage occurred in the office scene, a very bright and uniformly colored scene relative to the other scenes.
A one-way repeated ANOVA analysis showed that the shading condition ($\conditionOurs$ vs. $\conditionUniform$) has a significant effect on the perceptual quality ($F_{(1, 24)}=18.42, p=.00025$). 

As plotted in \Cref{fig:eval:bar-power}, we also measured the display power consumption of each power-saved shading condition for each scene. The average savings between $\conditionOurs$ and $\conditionUniform$ are similar ($20.8\pm 1.2\%$ vs. $18.6\pm 1.4\%$ (95\% confidence)). 
$\conditionOurs$ exhibits the highest power saving in the skyline scene, due to higher relative uniform distribution of blue colors.

\paragraph{Discussion}
The results reveal our method $\conditionOurs$'s significant out-performance on preserving perceptual quality over a gaze-contingent luminance-reduction-based approach ($\conditionUniform$), even though both conditions achieve a comparably similar power-saving scale. Note that, under the same power, there are infinite ways of constructing $\conditionUniform$, including smoothing the edge but darkening the farther periphery. Our implementation of $\conditionUniform$ is partially inspired by P{\"o}ppel and Harvey~\shortcite{poppel1973light}, which suggests that human luminance change detection thresholds remain relatively constant beyond $10^\circ$ eccentricity. The design, however, may not be perceptually optimal. Therefore, studying and modeling the luminance-induced effects may not only provide a stronger baseline condition, but improve our model that is currently restricted to colors only.

Our perceptual fidelity and power-saving capabilities are also content-based.
For example, we notice high power savings in the ``office'' scene, but the average \% of observed artifacts is higher than other scenes.
This is hypothetically because the scene has a significantly higher brightness compared to the others.
On the other hand, the ``monkey'' scene has relatively high density of green colors, and thus has the lowest perceived average \% of artifacts.
The observations motivate us to investigate, in the future, the chromaticity-luminance joint effect (\Cref{sec:future-work}) beyond the first model that guides color-perception-aware VR power optimization.

While detection tasks are commonly leveraged in the foveated rendering literature \cite{Patney:2016:TFR, Sun:2017:PGF}, we opted to validate our gaze contingent filter with active and natural viewing, similar to \cite{Cohen:2020:TLC}. 
The design choice is two-fold. 
First, we attempt to simulate the conditions of real-world VR applications where the users, with various tasks in mind, make head and eye movements to explore the environment.
Examples include gaming and video-watching.
Second, prior literature suggests unequal color detection and discrimination thresholds. 
Vingrys and Mahon~\shortcite{vingrys1998color} discovered that chromatic sensitivity for detection is significantly greater than for discrimination. 
However, by leveraging our model and shader in this experiment, we verify our hypothesis that our color sensitivity during natural and active vision is, in fact, lower than that of discrimination, and thus enable the method's applicability for broad VR scenarios.

\subsection{Measuring Power-Saving Capability for Broad Content}
\label{sec:result:imagenet}

\begin{figure*}
    \centering
    \subfloat[original]{
      \includegraphics[width=0.32\linewidth,valign=t]{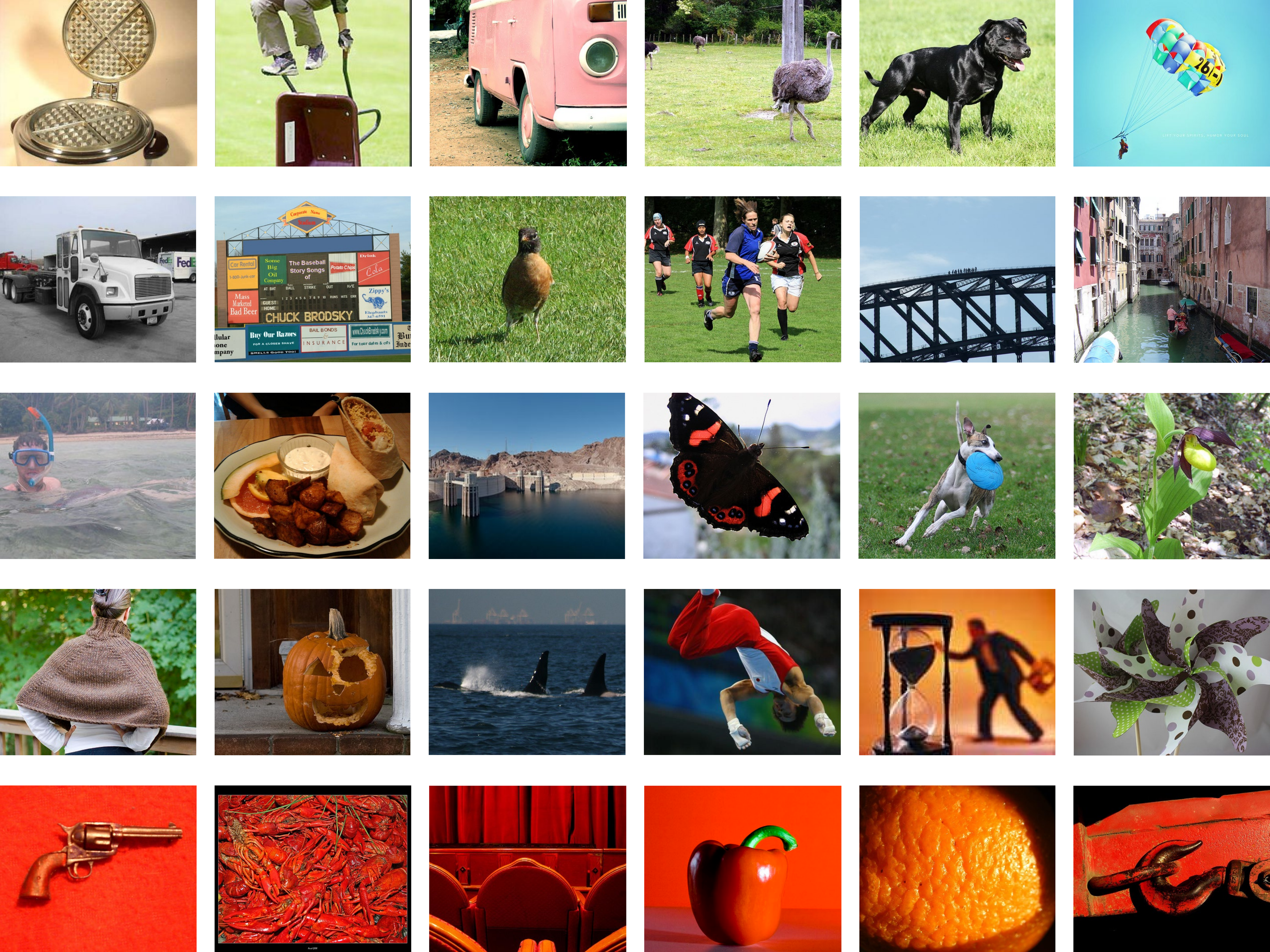}
      \vphantom{\includegraphics[width=0.32\linewidth,valign=t]{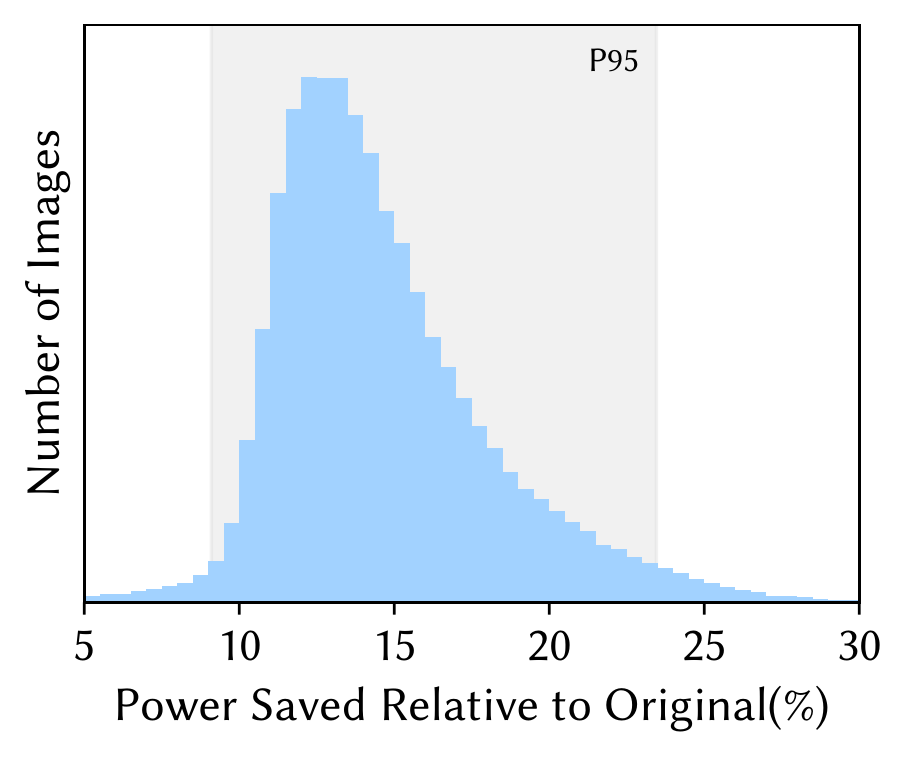}}
      \label{fig:result:imagenet-sample-gt}
    }%
    \subfloat[ours]{
      \includegraphics[width=0.32\linewidth,valign=t]{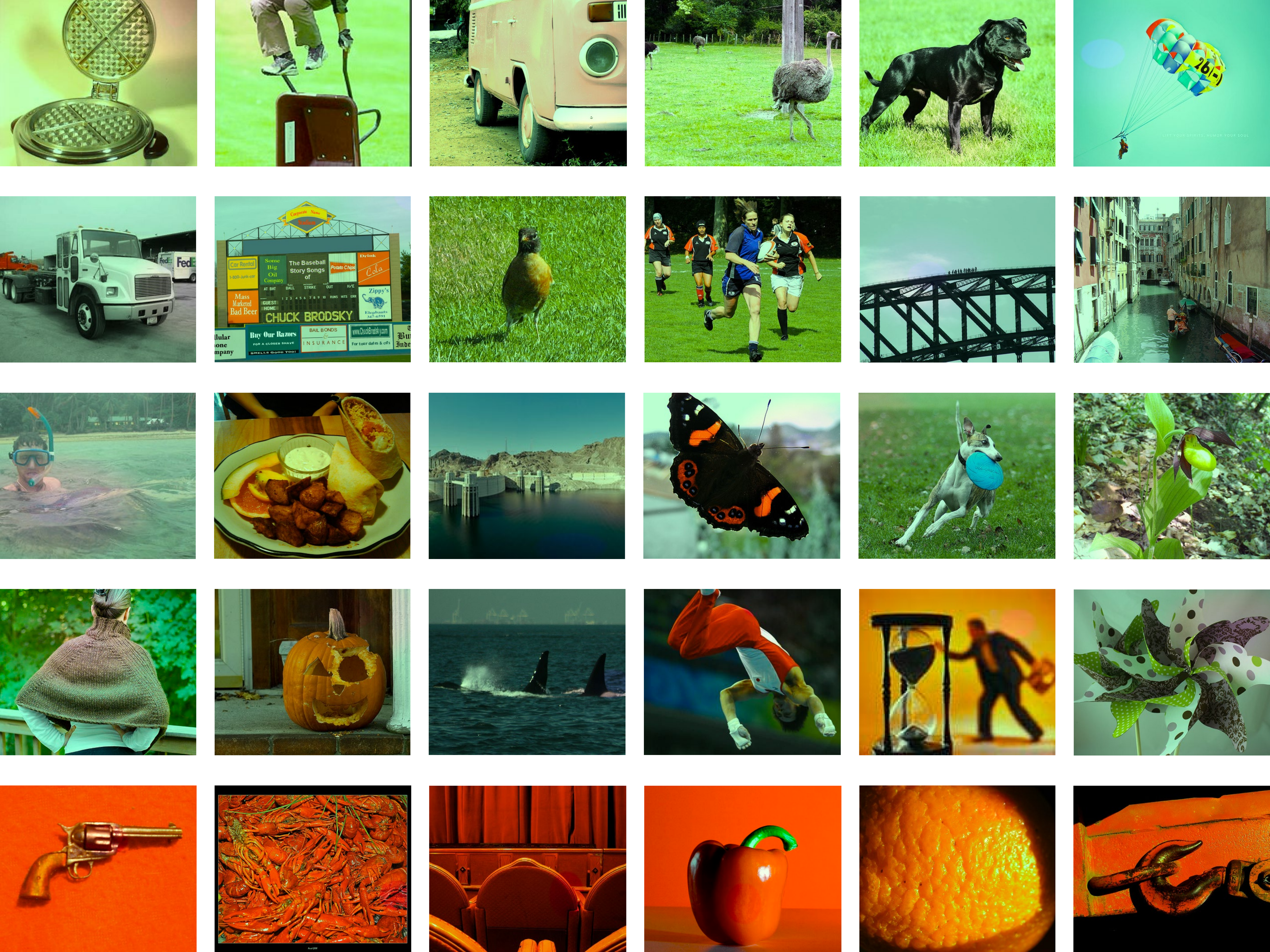}
      \vphantom{\includegraphics[width=0.32\linewidth,valign=t]{figures/results-imagenet/imagenet_histogram.pdf}}
      \label{fig:result:imagenet-sample-filtered}
    }%
    \subfloat[power savings histogram]{
      \includegraphics[width=0.32\linewidth,valign=t]{figures/results-imagenet/imagenet_histogram.pdf}
      \label{fig:result:imagenet-histogram}
    }%
    \Caption{
        ImageNet Offline Power Savings Estimation.
    }{
        We measure the statistics of power saved when our model is applied to the large database of natural images, ImageNet.
        \subref{fig:result:imagenet-sample-gt} The original images are sampled randomly to show exemplar images which exhibit different amounts of power-saving when our model is applied to them.
        Each row of images are sampled from within different quantile bands of energy savings across the dataset
        (top to bottom each row saves better than $99.9\%$, $55\%$, $45\%$, $0.1\%$ of the entire dataset considered, and the bottom row shows the worst $0.1\%$ performers).
        \subref{fig:result:imagenet-sample-filtered} Our shaders are applied to each image have randomized gaze locations to prevent bias against the edges of the image being disproportionately affected.
        \subref{fig:result:imagenet-histogram} We visualize the entire distribution of the potential power savings in this evaluation.
    }
    \Description{ImageNet Offline Power Savings Estimation.}
    \label{fig:result:imagenet}
\end{figure*}

In our psychophysical study \Cref{sec:result:perceptual}, we observe that the possible power savings are dependent on the displayed content.
For example, colors that are highly saturated have little room in their equi-luminant plane that is within the bounds of the \emph{sRGB} cube.
Therefore, they have less power saving potential as any potential power-saving chromaticity shifts are clipped by the \emph{sRGB} bounds.
To study how much power can be saved in practical applications where users may observe arbitrary imagery, we conduct an objective evaluation by applying our method to a large sample of the ImageNet dataset \cite{russakovsky2015imagenet}.
We then measure the distribution of power savings using our power model.

\paragraph{Setup}
We simulate how an image would be observed in a VR setting by resizing the image to be displayed at $90^\circ$ field-of-view, and randomly sample a location within the image and select that as the gaze location.
The randomization of the gaze location was applied to prevent bias in the power estimation.
Specifically, we observed that many images in ImageNet have a foreground object centered on the image; selecting random gaze locations allows us to include images where the foreground objects will sometimes have the filter applied to them.
We repeat this process for $10\%$ (randomly sampled) of the ImageNet dataset, totaling in over $\sim 120k$ images, to collect original and power-optimized image pairs.

\paragraph{Metrics}
For each image pair, we measure the estimated power consumption using our model from \Cref{sec:method:energy}, and compute the relative decrease in power consumption by applying our filter with respect to the ground-truth condition.

\paragraph{Results}
We observe that the mean display power saving recorded across the entire dataset is $13.9\%$, and guarantee $9.1 - 23.5\%$ savings with \emph{P95} confidence.
Please refer to \Cref{fig:result:imagenet-histogram} for the detailed histogram of the estimated power savings.
We visualize a small sample of the images we applied the filter to in \Cref{fig:result:imagenet-sample-gt,fig:result:imagenet-sample-filtered}.

\paragraph{Discussion}
Sample images from different percentiles of power-saving as shown in \Cref{fig:result:imagenet} show that images with the highest power savings are commonly bright and/or blueish scenes, and vice versa.
Intuitively, bright scenes provide larger \emph{percentage} changes in LED luminance, and thus unlock larger space for power-saving.
Since blue colors on the LED consume the most power, as demonstrated in \Cref{sec:display-energy-study}, images rich in blue/green colors can be optimized most effectively.
Meanwhile, images which are already saturated with red colors cannot be optimized for higher power-saving because the space of power-wise ``cheaper'' colors is narrower.

\section{Limitations and Future Work}
\label{sec:future-work}

\paragraph{Active vision vs. discrimination vs. detection.} While our evaluation on active/natural viewing tasks in \Cref{sec:result:perceptual} is representative of real-world VR scenarios \cite{Cohen:2020:TLC}, our initial perceptual data are collected using a more conservative \textit{discrimination} task.
It is also common in the foveated rendering literature to evaluate using \textit{detection} tasks \cite{Patney:2016:TFR,Sun:2017:PGF}.
Our conducted preliminary detection-based experiments, which showed that sensitivity to color changes in detection tasks is significantly greater that that in discrimination tasks, are consistent with prior work \cite{vingrys1998color}.

An exciting future direction is, thus, to investigate an adaptive model that accommodates for color sensitivity under all three tasks (detection, discrimination, active/natural viewing). That way, our color modulation algorithm can be dynamically configured according to the specific viewing scenario of a VR user.

\paragraph{Perceptual model.} Our current perceptual model is constructed with respect to per-pixel colors. An interesting future extension is to consider inter-pixel, potentially higher-dimensional, features such as spatial frequency and local contrast. Performing these analyses (e.g., frequency domain analyses as in \cite{tursun2019luminance}), however, increases the computational overhead. How to best balance the level of details in perceptual analysis and display power saving is an open question we leave to future work.

\paragraph{Luminance adjustment.} In our work, we model and modulate pixel \textit{chromaticity} to reduce display power consumption while preserving \textit{luminance}.
This design choice reduces the dimension of our perceptual model and, thus, yields a convex constrained optimization problem with a closed-form solution.
Investigating the luminance-chromaticity joint modulation is an interesting future research direction that would conceivably lead to higher power savings \cite{vingrys1998color}.

Jointly adjusting luminance and chromaticity, nevertheless, comes with a few challenges. First, it would require sampling a new dimension in constructing the perceptual model.
Second, prior literature suggests the weak eccentricity-dependent effect \cite{metha1994detection} in detecting and discriminating absolute luminance. Finally, the perceptual level sets, when considering the luminance dimension, might not be convex, which might complicate the optimization, cause false local minima, and reduce the shading speed.

\paragraph{Color Temperature Adaptation.} Another interesting direction for future research is to leverage chromatic adaptation~\cite{fairchild2013color} to reduce display power by adjusting the color temperature of the display white point. The advantage of this approach is that it is not gaze-contingent: it can potentially reduce the power of the entire display without requiring eye tracking.
Adaptation to display color has been long investigated \cite{fairchild1995time, peng2021white}, but such studies in VR displays are relatively new and rare~\cite{chinazzo2021temperature} and lack a comprehensive computational model. Note that the chromatic adaptation benefits are additive: our model can be seen as a sub-space initial attempt (by exploiting spatial color perception) under a \textit{given} adaptation state.

\paragraph{Display Persistence.}
VR displays usually have low persistence to reduce motion blur~\cite{hainich2016displays}.
A common solution is to hold a frame for only a short period of time during each display refresh \cite{Google:2019:VR}.
As a consequence, the display power is relatively low to begin with (compared to displaying a frame throughout a refresh cycle).
Nevertheless, our work demonstrates significant display power saving opportunities even with reduced displayed times.
In addition, reducing the display period leads to low average luminance, which limits the applicability of a luminance-based approach to reduce power --- another reason we choose to maintain the luminance.

\paragraph{Implementation.}
Our implementation of \emph{LMS} to \emph{DKL} color space transformation does not strictly follow the canonical implementation \cite{Derrington:1984:DKL, westland2012computational}. Notably, we judiciously choose to use a per-luminance adaptation color rather than one single adaptation color. Effectively, this allows us to eliminate the luminance dimension from all colors. As a result, all subsequent color adjustments naturally preserve the luminance --- a goal we set out to achieve. This design choice also simplifies the color space transformation and contributes to the real-time speed of the shader.

There is room for improving the speed of our shader, which currently is bottlenecked by the atomic \texttt{for}-loop. Deferred shading techniques may shed light on alleviating the bottleneck. One promising solution is to evaluate the optimization problem (\Cref{eq:method:power:opt}) offline (e.g., sampling colors and eccentricies) and save the results as a 3D texture, which the shader simply looks up at rendering time.

Due to the tight integration of the display, computation module, and battery in commercial AR/VR devices, our display power measurement has to be done on a 3rd party display module that has the identical aspect ratio of the VR device we use for perceptual studies. Investigating physical means to measure the exact display power as in an AR/VR device would reveal the real-world energy savings concerning the battery equipped with the device.
It would also be interesting to see how our perception-conserving color modulation idea can be applied to smartphone displays, which have much narrower field-of-views.

\section{Conclusion}
\label{sec:conclusion}

Understanding the nature of color generation and perception is a long-standing pillar for computer graphics. In this research, we build a bridge from color to the emerging demand for power-friendly graphics with the rapid growth of wireless platforms. 
In particular, we present a perceptually motivated computational model for optimizing the power usage of VR displays based on the limited discriminative ability of color by the human visual system. 
We validate the model's application with various real-world immersive content viewing use-cases and physically-measured power consumptions.
While the research is now benefiting untethered VR applications, we envision potential in exploiting more properties of the ``human'' aspects of the study of color and how to incorporate them into generic computer graphics systems. We hope that our work to develop the first building block in this direction will pave the way for more sophisticated applications, such as cloud rendering.

\begin{acks}

The work was supported, in part, by the National Science Foundation (NSF) under grants \#2225861 and \#2044963.

\end{acks}

\bibliographystyle{acmart}
\bibliography{paper}

\appendixpageoff
\appendixtitleoff
\renewcommand{\appendixtocname}{Supplementary material}
\begin{appendices}

\crefalias{section}{supp}
\normalsize

\include{\jobname-support}

\end{appendices}

\end{document}